\documentclass[draftcls, onecolumn]{IEEEtran}
\ifCLASSINFOpdf
\else
\fi
\usepackage{graphicx}
\usepackage{amsmath}
\usepackage{amssymb}
\usepackage{graphics}
\usepackage{latexsym}
\usepackage[dvips]{epsfig}
\usepackage[nospace]{cite}
\usepackage{subfigure}
\usepackage{setspace}
\usepackage{tabularx}
\usepackage{color}

\newtheorem{Definition}{Definition}
\newtheorem{Lemma}{Lemma}
\newtheorem{Corollary}[Lemma]{Corollary}

\newtheorem{Theorem}{Theorem}

\newtheorem{Remark}{Remark}

\usepackage[ colorlinks = true,
 linkcolor = blue,
 urlcolor = blue,
 citecolor = red,
 anchorcolor = green,
]{hyperref}

\allowdisplaybreaks[1]

\def\Pr{{\mathbb{P}}}
\def\E{{\mathbb{E}}}

\begin{document}
%
\title{Non-Asymptotic Achievable Rates for Gaussian Energy-Harvesting Channels: \\Save-and-Transmit and Best-Effort}

\author{Silas L.~Fong,
Jing Yang, and Aylin Yener
\thanks{This work was sponsored by NSF under Grant CCF-1422347, Grant CNS-1526165, Grant ECCS-1650299, and Grant ECCS-1748725.}
\thanks{This paper was presented in part at the 2018 IEEE International Symposium on Information Theory.}
\thanks{S.~L.~Fong is with the Department of Electrical and Computer Engineering, University of Toronto, Toronto, ON M5S 3G4, Canada  (e-mail: \texttt{silas.fong@utoronto.ca}).}
\thanks{J.~Yang and A.~Yener are with the Department of Electrical Engineering, The Pennsylvania State University, University Park, PA 16802, USA (e-mail:
\texttt{\{yangjing,yener\}@engr.psu.edu}).}

}


%


\maketitle

\begin{abstract}
An additive white Gaussian noise energy-harvesting channel with an infinite-sized battery is considered. The energy arrival process is modeled as a sequence of independent and identically distributed random variables. The channel capacity $\frac{1}{2}\log(1+P)$ is achievable by the so-called best-effort and save-and-transmit schemes where $P$ denotes the battery recharge rate. This paper analyzes the save-and-transmit scheme whose transmit power is strictly less than $P$ and the best-effort scheme as a special case of save-and-transmit without a saving phase. In the finite blocklength regime, we obtain new non-asymptotic achievable rates for these schemes that approach the capacity with gaps vanishing at rates proportional to $1/\sqrt{n}$ and $\sqrt{(\log n)/n}$ respectively where~$n$ denotes the blocklength. The proof technique involves analyzing the escape probability of a Markov process. When $P$ is sufficiently large, we show that allowing the transmit power to back off from $P$ can improve the performance for save-and-transmit. The results are extended to a block energy arrival model where the length of each energy block $L$ grows sublinearly in $n$. We show that the save-and-transmit and best-effort schemes achieve coding rates that approach the capacity with gaps vanishing at rates proportional to $\sqrt{L/n}$ and $\sqrt{\max\{\log n, L\}/n}$ respectively.
\end{abstract}


%
\IEEEpeerreviewmaketitle

\section{Introduction} \label{Introduction}
In this paper, we consider communication over an energy-harvesting (EH) channel which has an input alphabet~$\mathcal{X}$, an output alphabet~$\mathcal{Y}$ and an infinite-sized battery that stores energy harvested from the environment. The channel law of the EH channel is characterized by a conditional distribution $q_{Y|X}$ where $X\in\mathcal{X}$ and $Y\in\mathcal{Y}$ denote the channel input and output respectively. A source node wants to transmit a message to a destination node through the EH channel. Let $\mathrm{c}: \mathcal{X}\rightarrow [0, \infty)$ be a cost function associated with the EH channel, where $\mathrm{c}(x)$ represents the amount of energy used for transmitting $x\in\mathcal{X}$. At each discrete time $k \in \{1,2,\ldots\}$, a random amount of energy $E_k$ arrives at the battery buffer and the source transmits a symbol $X_k\in \mathcal{X}$ such that
 \begin{equation}
 \sum_{i=1}^k \mathrm{c}(X_i) \le \sum_{i=1}^k E_i   \qquad\text{almost surely.}  \label{EHResultIntro}
 \end{equation}
 This implies that the total harvested energy $\sum_{i=1}^k E_i$ must be no smaller than the ``energy" of the codeword $\sum_{i=1}^k \mathrm{c}(X_i)$  at every discrete time~$k$ for transmission to take place successfully. The destination receives $Y_k$ from the channel output in time slot~$k$ for each $k\in\{1, 2, \ldots\}$, where $(X_k, Y_k)$ is distributed according to the channel law such that $p_{Y_k|X_k}(y_k|x_k)=q_{Y|X}(y_k|x_k)$ for all $(x_k, y_k)\in \mathcal{X}\times \mathcal{Y}$.
We assume that $\{E_i\}_{i=1}^\infty$ are independent and identically distributed (i.i.d.), where $E_1$ is a non-negative random variable. To simplify notation, we write $E\triangleq E_1$ if there is no ambiguity. Throughout the paper, we let $P\triangleq\E[E]$, the expected value of~$E$, denote the battery recharge rate, and we assume that $\E[E^2]<\infty$. All results presented in this paper depend on the random variable~$E$ only through its first and second moments rather than its distribution.

This paper focuses on the additive white Gaussian noise (AWGN) model where $\mathcal{X}=\mathcal{Y}=\mathbb{R}$, $q_{Y|X}(y|x)\equiv\frac{1}{\sqrt{2\pi}}\,\mathrm{e}^{-\frac{(y-x)^2}{2}}$ and $\mathrm{c}(x)\equiv x^2$. Under the AWGN model, the received symbol at time~$k$ can be expressed as
\begin{align}
Y_k=X_k+Z_k \label{defChannelLawIntro}
\end{align}
for each time~$k$ where $Z_k$ is a standard normal random variable which is independent of~$X_k$ and the random variables $\{Z_k\}_{k=1}^\infty$ are independent. Reference~\cite{ozel12} has shown that the capacity of this channel is
$\frac{1}{2}\log(1+P)$ and proposed two capacity-achieving schemes, namely \emph{save-and-transmit} and \emph{best-effort}.

The save-and-transmit scheme consists of an initial saving phase and a subsequent transmission phase. The transmitter remains silent in the saving phase so that energy accumulates in the battery. In the transmission phase, the transmitter sends the symbols of a random Gaussian codeword with variance~$P-\nu$ as long as the battery has sufficient energy where $\nu \in[0,P)$ denotes some small offset from~$P$.

The best-effort scheme has a simpler design than the save-and-transmit scheme as it does not have an initial saving phase. As long as the transmitter has sufficient energy to output the symbols of a random Gaussian codeword with variance~$P-\nu$ for some $\nu \in[0,P)$, information gets transmitted.

Following reference~\cite{ozel12}, a number of non-asymptotic achievable rates for the save-and-transmit scheme have been presented in references~\cite{FTY15,ShenoySharma16,FTO17}. By contrast, no non-asymptotic achievable rate exists for the best-effort scheme except for a special discrete memoryless EH channel with infinite battery studied in~\cite{yangEH2014} and a special discrete memoryless EH channel with no battery studied in~\cite{MolavianJaziYener2015}. A main goal of this paper is to provide a non-asymptotic achievable rate for save-and-transmit with a saving phase of arbitrary length, which will immediately imply a non-asymptotic achievable rate for best-effort.

Note that the results in this paper cease to hold if the size of the battery is finite. The channel capacity for the finite battery case is the subject of recent interests, see~\cite{TOYU2017,SNOzgur2016,MaoHassibi2017}.
\subsection{Related Work} \label{subsecRelatedWork}
The channel capacity of the AWGN EH channel was characterized in~\cite{ozel12}, which showed that the capacity of the AWGN channel with an infinite-sized battery subject to EH constraints is equal to the capacity of the same channel under an average power constraint where the average power equals the average recharge rate of the battery. In particular, \emph{save-and-transmit}~\cite[Sec.~IV]{ozel12} and \emph{best-effort}~\cite[Sec.~V]{ozel12} were proposed as capacity-achieving strategies.

For a fixed tolerable error probability~$\varepsilon$, reference.~\cite{FTY15} has performed a finite blocklength analysis of save-and-transmit proposed in~\cite{ozel12} and obtained a non-asymptotic achievable rate for the AWGN EH channel. The first-, second- and third-order terms of the non-asymptotic achievable rate presented in~\cite[Th.~1]{FTY15} are equal to the capacity, $-O\Big(\sqrt{\frac{\log n}{n}}\Big)$ and $-O\Big(\sqrt{\frac{2+\varepsilon}{n \varepsilon}}\Big)$ respectively where the big-$O$ notation is used for a positive term which involves the blocklength~$n$ and which approaches zero at a rate no slower than the argument of the notation as~$n$ approaches infinity. The formal definition of the big-$O$ notation can be found in Section~\ref{notation}. Subsequently, reference~\cite{ShenoySharma16} has refined the analysis in~\cite{FTY15} and improved the second-order term to $-O(1/\sqrt{n \varepsilon})$.
Reference~\cite{FTO17} has further improved the second-order term to $\sup\limits_{\substack{\varepsilon_1\ge 0, \varepsilon_2\ge 0,\\ \varepsilon_1+\varepsilon_2=\varepsilon}}-O\Big(\sqrt{\frac{\log(1/\varepsilon_2)}{n}}\Big)+\sqrt{ \frac{P}{n(P+1)}}\,\Phi^{-1}(\varepsilon_1)$ if $\varepsilon\in (0, 1/2)$ where $\Phi(\cdot)$ denotes the cumulative density function (cdf) of the standard normal random variable. All the second-order terms obtained by the above studies and the current study are inferior (more negative) to the following second-order term corresponding to the non-EH AWGN channel where all energy is available to the transmitter at the onset and~\eqref{EHResultIntro} is replaced with the conventional power constraint $\Pr\{\frac{1}{n}\sum_{k=1}^n X_k^2 \le nP\}=1$ \cite[Th.~54]{Pol10}: $\sqrt{\frac{P(P+2)}{2n(P+1)^2}}\Phi^{-1}(\varepsilon)$.

For the block energy arrival model where the length of each energy block~$L$ grows sublinearly in~$n$~\cite{ZhangLau14,shavivOzgur18,FTO17}, reference~\cite{FTO17} has proved that save-and-transmit achieves the second-order term $-O\Big(\sqrt{\frac{\log(1/\varepsilon)L}{n}}\Big)$ if $\varepsilon\in (0, 1/2)$. In addition, a non-asymptotic upper bound~$\frac{\sqrt{2P^2+\E[E^2]}}{2(P+1)}\Phi^{-1}(\varepsilon)\times \sqrt{\frac{L}{n}}$ on the second-order term has been proved in~\cite{FTO17} for a general coding scheme, implying that save-and-transmit achieves the optimal second-order scaling $-O(\sqrt{L/n})$ if $\varepsilon\in (0, 1/2)$.

\subsection{Main Contributions}
In this paper, we analyze a save-and-transmit scheme with a saving phase of arbitrary length (including zero, which corresponds to the best-effort scheme) and derive a non-asymptotic achievable rate. The derivation involves designing the transmit power to be strictly less than the battery recharge rate~$P$ (unlike the design in~\cite{FTY15,FTO17} which sets the transmit power equal to~$P$) so that we can effectively bound the number of mismatched positions between the desired transmitted codeword and the actual transmitted codeword subject to a fixed blocklength.
The aforementioned non-asymptotic achievable rate is extended to the block energy arrival model where the length of each energy block~$L$ grows sublinearly in~$n$~\cite{ZhangLau14,shavivOzgur18,FTO17}. Our analyzed best-effort and save-and-transmit achieve the second-order scalings $-O\Big(\sqrt{\frac{\max\{\log n, L\}}{n}}\Big)$ and $-O(\sqrt{L/n})$ respectively. If $\varepsilon\in(0,1/2)$, the second-order term for a general coding scheme has been proved to be bounded above by $-O(\sqrt{L/n})$ as explained in the previous subsection, which implies that both analyzed schemes achieve the optimal second-order scaling~$-O(\sqrt{L/n})$ if $L$ grows faster than~$\log n$.

In order to compare our results with the existing ones, we focus on the i.i.d.\ energy arrival case (i.e., $L=1$) in the remainder of this subsection. This work provides the first finite blocklength analysis of the best-effort scheme for the AWGN EH channel and presents a non-asymptotic achievable rate. It shows that the first- and second-order terms of the asymptotic achievable rate are equal to the capacity and $-O\Big(\sqrt{\frac{\log(1/\varepsilon)\log n}{n}}\Big)$ respectively. This second-order scaling $-O\Big(\sqrt{\frac{\log(1/\varepsilon)\log n}{n}}\Big)$ significantly improves the state-of-the-art result in~\cite{ozel12} which did not derive a bound on the vanishing rate for the second-order term.
In addition, this work obtains a new non-asymptotic achievable rate for save-and-transmit which outperforms the state-of-the-art result for save-and-transmit with transmit power equal to~$P$~\cite[Th.~1]{FTO17}.
%
%
\subsection{Paper Outline}
This paper is organized as follows. The notation of this paper is presented in the next subsection. Section~\ref{sectionDefinitionAWGN} presents the model of the AWGN EH channel. Section~\ref{sectionSaveTransmit} describes the save-and-transmit scheme, states the corresponding preliminary results, and presents the main result --- a new non-asymptotic achievable rate for save-and-transmit with a saving phase of arbitrary length. A non-asymptotic achievable rate for best-effort is then obtained by setting the length of the saving phase to be zero. Section~\ref{sectionBlockEH} generalizes the non-asymptotic results in Section~\ref{sectionSaveTransmit} to the block energy arrival model. Section~\ref{sectionThmMainResultBlock} presents the proof of the new non-asymptotic achievable rate for save-and-transmit for the block energy arrival model which subsumes the proof for the i.i.d.\ energy arrival model.
Section~\ref{sectionNumerical} contains numerical results which demonstrate the performance advantage of allowing the transmit power for a save-and-transmit to back off from the battery recharge rate in the high battery recharge rate regime for both i.i.d.\ and block energy arrivals. Section~\ref{conclusion} concludes the paper.

\subsection{Notation}\label{notation}
We use $O(\cdot)$, $\Theta(\cdot)$, $\omega(\cdot)$ and $o(\cdot)$ to denote standard asymptotic Bachmann-Landau notations that involve the blocklength variable~$n$ except our convention that they must be positive. Therefore, we have $\limsup\limits_{n\to \infty}\frac{O(\cdot)}{n}<\infty$, $0<\liminf\limits_{n\to \infty} \frac{\Theta(\cdot)}{n}\le \limsup\limits_{n\to \infty}\frac{\Theta(\cdot)}{n}<\infty$, $\limsup\limits_{n\to \infty}\frac{\omega(\cdot)}{n}=\infty$, and  $\lim\limits_{n\to \infty}\frac{o(\cdot)}{n}=0$.
The sets of natural numbers, real numbers and non-negative real numbers are denoted by $\mathbb{N}$, $\mathbb{R}$ and $\mathbb{R}_+$ respectively.
All logarithms are taken to base~$\mathrm{e}$ throughout the paper.

We use $\Pr\{\mathcal{E}\}$ to represent the probability of an
event~$\mathcal{E}$, and we let $\mathbf{1}\{\mathcal{E}\}$ be the indicator function of $\mathcal{E}$. Random variables are denoted by capital letters (e.g., $X$), and the realization and the alphabet of a random variable are denoted by the corresponding small letter (e.g., $x$) and calligraphic font (e.g., $\mathcal{X}$) respectively.
We use $X^n$ to denote a random tuple $(X_1,  X_2,  \ldots ,  X_n)$, where all of the elements $X_k$ have the same alphabet~$\mathcal{X}$.
We let $p_X$ and $p_{Y|X}$ denote the probability distribution of $X$ and the conditional probability distribution of $Y$ given $X$ respectively for random variables~$X$ and~$Y$.
We let $p_Xp_{Y|X}$ denote the joint distribution of $(X,Y)$, i.e., $p_Xp_{Y|X}(x,y)=p_X(x)p_{Y|X}(y|x)$ for all $x$ and $y$.
 For random variable~$X\sim p_X$ and any real-valued function~$g$ whose domain includes $\mathcal{X}$, we let $\Pr_{p_X}\{g(X)\ge\xi\}$ denote $\int_{\mathcal{X}} p_X(x)\mathbf{1}\{g(x)\ge\xi\}\, \mathrm{d}x$ for any real constant $\xi$.
For any function $f$ whose domain contains $\mathcal{X}$, we use $\E_{p_X}[f(X)]$ to denote the expectation of~$f(X)$ where $X$ is distributed according to $p_X$.
 For simplicity, we omit the subscript of a notation when there is no ambiguity.
The Euclidean norm of a tuple $a^L\in \mathbb{R}^L$ is denoted by $\|a^L\|\stackrel{\text{def}}{=} \sqrt{\sum_{\ell=1}^L a_\ell^2}$\,. The distribution of a Gaussian random variable~$Z$ whose mean and variance are $\mu$ and $\sigma^2$ respectively is denoted by $ \mathcal{N}(z; \mu, \sigma^2)\triangleq \frac{1}{\sqrt{2\pi \sigma^2}}\mathrm{e}^{-\frac{(z-\mu)^2}{2\sigma^2}}$.

\section{The AWGN EH Channel} \label{sectionDefinitionAWGN}
\subsection{Problem Formulation}
\begin{figure}[!t]
\centering
\includegraphics[width=4.4in]{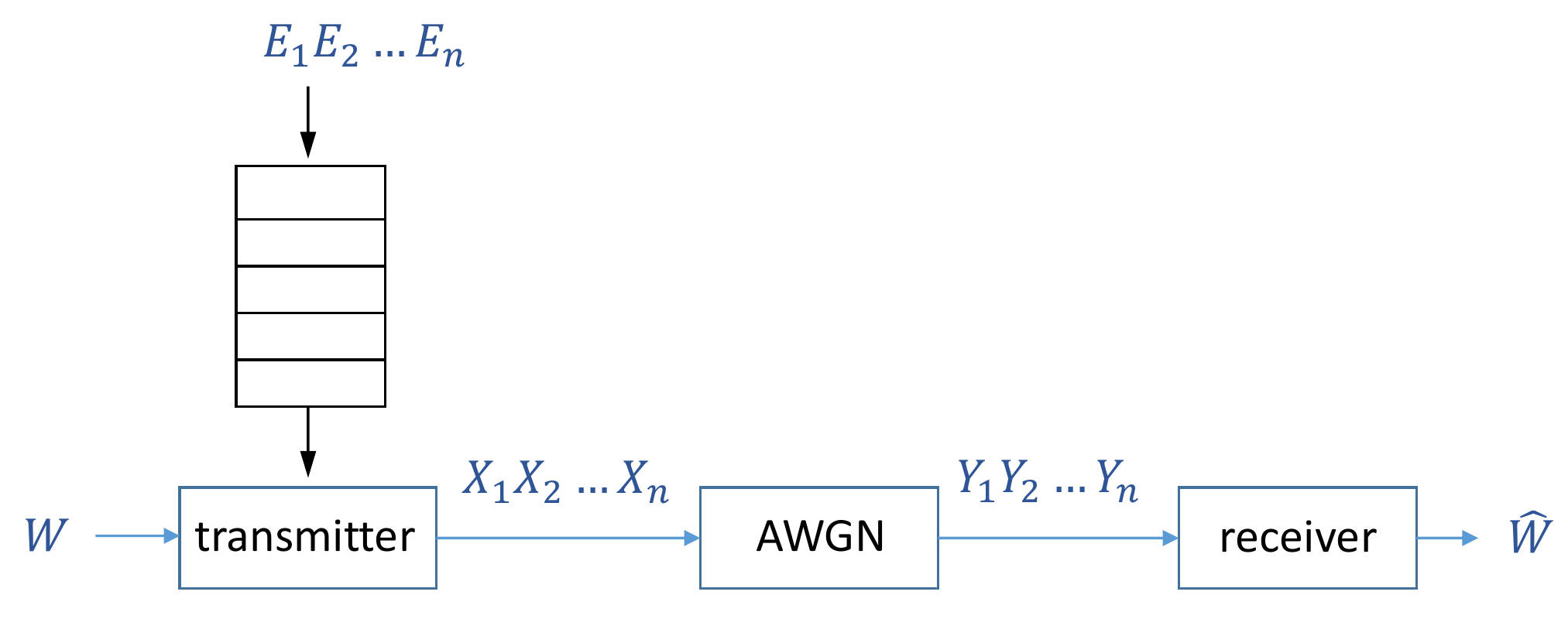}
 \vspace{-0.1 in}
 \caption{The AWGN EH channel}
 \vspace{-0.2 in}
\label{figure0}
\end{figure}
The AWGN EH channel, as illustrated in Figure~\ref{figure0}, consists of one transmitter and one receiver. Energy harvesting and communication occur in~$n$ time slots, i.e., channel uses. In each time slot, a random amount of energy~$E$ with alphabet~$\mathbb{R}_+$ is harvested where
\begin{equation}
0 < P = \E[E] \text{\qquad and \qquad} \E[E^2]<\infty. \label{defEnergyP}
\end{equation}
  The energy-harvesting process is characterized by~$n$ independent copies of~$E$ denoted by $E_1, E_2, \ldots, E_n$. Prior to communication, the transmitter chooses a message~$W$.
For each $k\in \{1, 2, \ldots, n\}$, the transmitter consumes $X_k^2$ units of energy to transmit $X_k\in \mathbb{R}$ based on~$(W, E^k)$ and the receiver observes $Y_k\in \mathbb{R}$ in time slot~$k$. The energy state information~$E_k$ is known by the transmitter at time~$k$ before encoding~$X_k$, but the receiver has no access to~$E_k$.
For each $k\in\{1, 2, \ldots, n\}$, we have:
\begin{enumerate}
\item[(i)] $E_k$ and $(W, E^{k-1}, X^{k-1}, Y^{k-1})$ are independent, i.e.,
\begin{align}
p_{W, E^k, X^{k-1}, Y^{k-1}} = p_{E_k}p_{W, E^{k-1}, X^{k-1}, Y^{k-1}}. \label{assumption(i)}
\end{align}
    \item[(ii)] For $w\in\mathcal{W}$ and every $e^n\in\mathbb{R}_+^n$, a transmitted codeword $X^n$ should satisfy
\begin{equation}
\Pr\left\{\left.\sum_{i=1}^k X_i^2 \le \sum_{i=1}^k e_i \right|\, W=w, E^n = e^n\right\}=1 \label{eqn:eh}
\end{equation}
for each $k\in\{1, 2, \ldots, n\}$.
  \end{enumerate}
   After the~$n$ time slots, the receiver declares~$\hat W$ to be the transmitted~$W$ based on $Y^n$.
   \subsection{Standard Definitions}
   Formally, we define a code as follows:
\begin{Definition} \label{defCode}
An {\em $(n, M)$-code} consists of the following:
\begin{enumerate}
\item A message set
$
\mathcal{W}\triangleq \{1, 2, \ldots, M\}
$, where $W$ is uniform on $\mathcal{W}$.

\item A sequence of encoding functions
$
f_k : \mathcal{W}\times \mathbb{R}_+^k\rightarrow \mathbb{R}
$
 for each $k\in\{1, 2, \ldots, n\}$, where $f_k$ is used by the transmitter at time slot~$k$ for encoding $X_k$ according to $X_k=f_k (W, E^k)$.
\item A decoding function
$
\varphi :
\mathbb{R}^{n} \rightarrow \mathcal{W}
$
for decoding $W$ at the receiver, i.e., $\hat W = \varphi(Y^{n})$.
\end{enumerate}
If the sequence of encoding functions $f_i$ satisfies~\eqref{eqn:eh}, the code is also called an {\em $(n, M)$-EH code}.
\end{Definition}
\medskip

If an $(n, M)$-code does not satisfy the EH constraints~\eqref{eqn:eh} during the encoding process (i.e., $X^n$ is a function of~$W$ alone), then the $(n, M)$-EH code can be viewed as an $(n, M)$-code for the usual AWGN channel without any cost constraint~\cite{Cov06,elgamal}. The following definition is a formal statement of the channel law~\eqref{defChannelLawIntro}.
\medskip
\begin{Definition}\label{defChannel}
The {\em AWGN EH channel} is characterized by a conditional probability distribution $q_{Y|X}(y|x)\triangleq \mathcal{N}(y; x, 1)$ such that the following holds for any $(n, M)$-code: For each $k\in\{1, 2, \ldots, n\}$,
\begin{align*}
p_{W, E^k, X^k, Y^k}
 = p_{W, E^k, X^k, Y^{k-1}}p_{Y_k|X_k} 
\end{align*}
where
\begin{equation*}
p_{Y_k|X_k}(y_k|x_k) = q_{Y|X}(y_k|x_k)=\frac{1}{\sqrt{2\pi}}\,\mathrm{e}^{-\frac{(y_k-x_k)^2}{2}} 
\end{equation*}
for all $x_k\in \mathcal{X}$ and $y_k\in \mathcal{Y}$.
\end{Definition}
\medskip

 For any $(n, M)$-code defined on the AWGN EH channel, let $p_{W,E^n, X^n, Y^n, \hat W}$ be the joint distribution induced by the code. We can factorize $p_{W,E^n, X^n, Y^n, \hat W}$ as
\begin{align}
 p_{W,E^n, X^n, Y^n, \hat W}
=p_W \left(\prod_{k=1}^n p_{E_k} p_{X_k|W, E^k} p_{Y_k|X_k}\right)p_{\hat W |Y^n}, \label{memorylessStatement}
\end{align}
which follows from the i.i.d.\ assumption of the EH process $E^n$ in~\eqref{assumption(i)}, the fact that $X_i$ is a function of $(W, E^i)$ (cf.\ Definition~\ref{defCode}) and the memoryless property of the channel $q_{Y|X}$ described in Definition~\ref{defChannel}.
\smallskip
\begin{Definition} \label{defErrorProbability}
For an $(n, M)$-code defined on the AWGN EH channel, we can calculate according to~\eqref{memorylessStatement} the \textit{average probability of decoding error} defined as $\Pr\big\{\hat W \ne W\big\}$.
We call an $(n, M)$-EH code with average probability of decoding error no larger than~$\varepsilon$ an {\em $(n, M, \varepsilon)$-EH code}.
\end{Definition}
\smallskip
\begin{Definition} \label{defAchievableRate}
Let $\varepsilon\in (0,1)$ be a real number. A rate $R$ is said to be \textit{$\varepsilon$-achievable} for the EH channel if there exists a sequence of $(n, M_n, \varepsilon)$-EH codes such that
\begin{equation*}
\liminf_{n\rightarrow \infty}\frac{1}{n}\log M_n \ge R.
\end{equation*}
\end{Definition}

\begin{Definition}\label{defCapacity}
The {\em $\varepsilon$-capacity} of the AWGN EH channel, denoted by $C_\varepsilon$, is defined to be
$
C_\varepsilon \triangleq \sup\{R: R\text{ is $\varepsilon$-achievable for the EH channel}\}$. The \emph{capacity} of the AWGN EH channel is $C\triangleq \inf_{\varepsilon>0}C_\varepsilon $.
\end{Definition}
\medskip

Define the capacity function
\begin{equation*}
\mathrm{C}(x) \triangleq
\frac{1}{2}\log(1+x) 
\end{equation*}
for all $x\ge 0$.
It was shown in~\cite[Sec.\ III]{ozel12} (see also \cite[Remark 1]{FTY15}) that
 \begin{equation*}
 C_\varepsilon = C = \mathrm{C}(P)
 \end{equation*}
 for all $\varepsilon\in (0,1)$ where
$P = \E[E] 
$
can be interpreted as the signal-to-noise ratio (SNR) of the AWGN EH channel.

\section{An Achievable Rate for Save-and-Transmit} \label{sectionSaveTransmit}
This section will present a non-asymptotic achievable rate for save-and-transmit. To this end, we first formally describe save-and-transmit in the following subsection.
 \subsection{Save-and-Transmit Scheme} \label{subsecSaveTransmit}
Fix a blocklength~$n$. Choose a positive real number $S<P=\E[E]$ that may depend on~$n$ and let
 \begin{align}
 p_X(x)\equiv \mathcal{N}(x;0, S) \label{defDistPX}
 \end{align}
such that $S= \E_{p_X}[X^2]$.
   The codebook consists of~$M$ mutually independent random codewords, which are constructed as follows. For each message~$w\in\mathcal{W}$, a length-$n$ codeword $X^n(w)\triangleq (X_1(w), X_2(w),\ldots X_n(w))$ consisting of~$n$ i.i.d.\ symbols is constructed where $X_1(w)\sim p_X$. In other words, the codebook consists of~$M$ i.i.d.\ Gaussian codewords where each codeword consists of~$n$ i.i.d.\ Gaussian random variables and has average power~$S$.

   Suppose $W=w$ and $E^n=e^n$, i.e., the transmitter chooses message~$w\in\mathcal{W}$ and the realization of $E^n$ is $e^n\in\mathbb{R}_+^n$. Then, the transmitter uses the following \emph{save-and-transmit $(n, M)$-EH code} with encoding functions~$\{f_k\}_{k=1}^n$ and decoding function~$\varphi$. The save-and-transmit code consists of an initial saving phase and a subsequent transmission phase. Define $m$ to be the number of time slots in the initial saving phase during which energy is harvested but not consumed and no information is conveyed. Define $f_1, f_2, \ldots, f_n$ in a recursive manner where
\begin{equation}
f_k(w, e^k)\triangleq
 \begin{cases}
 X_k(w) & \text{if $k>m$ and $\big(X_k(w)\big)^2 \le e_k + \sum\limits_{i=1}^{k-1} \left(e_i - \big(f_i(w, e^i)\big)^2 \right)$,}\\
 0 & \text{otherwise.}
\end{cases} \label{defSaveEncoding}
\end{equation}
For each $k\in\{1, 2, \ldots, n\}$, let $\tilde X_k(W)\triangleq f_k(W, E^k)$ be the symbol transmitted
at time~$k$.
By construction,
\begin{align*}
\Pr\left\{\left.\sum\limits_{i=1}^{k}\big(\tilde X_i(w)\big)^2 \le \sum\limits_{i=1}^{k} e_i \right|  W=w, E^n=e^n \right\}=1
\end{align*}
for each $k\in\{1, 2, \ldots, n\}$.
Upon receiving $\tilde Y^n (W)\triangleq (\tilde Y_1(W), \tilde Y_2(W), \ldots, \tilde Y_n(W))$ where $\tilde Y_k(W)$ is generated according to
\begin{equation}
\Pr\{\tilde Y_k(W) =b \,|\, \tilde X_k(W)=a \}\equiv q_{Y|X}(b|a), \label{channelLawDeltaAWGN}
\end{equation}
the receiver declares that $\varphi(\tilde Y^n(W))=j$ if $j$ is the unique integer in~$\mathcal{W}$ that satisfies
\begin{equation*}
\sum_{k=m+1}^n \log\frac{q_{Y|X}(\tilde Y_k(W)|X_k(j))}{p_{Y}(\tilde Y_k(W))} \ge \log \xi, 
\end{equation*}
where $p_Y$ is the marginal distribution of $p_X q_{Y|X}$ and $\log \xi$ is an arbitrary threshold to be carefully chosen later (cf.\ \eqref{defXiAWGNblock}).
Otherwise, the receiver chooses $\varphi(\tilde Y^n(W))\in\mathcal{W}$ according to the uniform distribution. The decoding is successful if $j=W$.

\subsection{Preliminaries}
An important quantity that determines the performance of the save-and-transmit $(n, M)$-EH code is
\begin{align}
\mathcal{Q}^{(n)}(w)\triangleq \left\{k\in\{m+1, m+2, \ldots, n\}\left|\tilde X_k(w) \ne X_k(w)\right.\right\}, \label{defSetQ}
\end{align}
which is a random set that specifies the mismatched positions between $\tilde X^n(w)$ and $X^n(w)$ during the transmission phase when the chosen message~$W$ equals~$w$. The following lemma presents an upper bound on the probability of seeing more than $\gamma+1$ mismatched positions in the transmission phase. The proof, which is based on analyzing the escape probability of a Markov process, is provided in Appendix~\ref{appendixAblock}.
\begin{Lemma} \label{lemmaSavingMismatch}
Fix any~$n$ and any $\rho\in(0, 1)$ such that
\begin{align}
\frac{\sqrt{42\rho}}{21} <\frac{\sqrt{1-\rho}}{2}\,, \label{sufficientlyLargeN}
\end{align}
 and fix a save-and-transmit $(n, M)$-EH code with a length-$m$ saving phase where
\begin{align}
S\triangleq  (1-\rho)P.\label{defS}
\end{align}
Define
  \begin{equation}
\alpha\triangleq \frac{2\rho P}{\E[E^2]+3S^2} \label{defAlpha}
 \end{equation}
 and
  \begin{equation}
\beta\triangleq \frac{\alpha}{1+63\alpha S}. \label{defBeta}
 \end{equation}
For any~$\gamma\in\mathbb{R}_+$, we have
\begin{align}
\Pr\left\{\left.|\mathcal{Q}^{(n)}(w)| \ge \gamma+1\right|W=w\right\} \le  \mathrm{e}^{-(m+\gamma)\big(P\beta+\frac{\alpha^2 \E[E^2]}{2}\big)} \label{stLemmaMismatch}
\end{align}
for each~$w\in\mathcal{W}$.
\end{Lemma}
\begin{Remark}
In the proof of Lemma~\ref{lemmaSavingMismatch} which is readily seen in Appendix~\ref{appendixAblock} by setting~$L=1$, $\hat X_i=X_i$ and $\hat E_i=E_i$, an important step is analyzing the escape probability~\eqref{appendixLemma1Eq0-Block} of the Markov process $\left\{\sum_{i=1}^mE_i+\sum_{i=m+1}^{m+k} \big(E_i-X_i^2\big)\right\}_{k=1}^\tau$ where $\tau$ is the stopping time when the value of the Markov process hits any negative number $a<0$.
\end{Remark}
\medskip

The following lemma~\cite{sha57} is standard for proving achievability results in the finite blocklength regime and its proof can be found in~\cite[Th.~3.8.1]{Han10}.
\begin{Lemma}[Implied by Shannon's bound~{\cite[Th.~1]{sha57}}] \label{lemmaFeinstein*}
Let $p_{X^n, Y^n}$ be the probability distribution of a pair of random variables $(X^n,Y^n)$. Suppose $(X^n(1), Y^n(1))\sim p_{X^n, Y^n}$, and suppose $X^n(2)$ has the same distribution as $X^n(1)$ and is independent of $Y^n(1)$.
Then for each $\delta>0$ and each~$M\in \mathbb{N}$, we have
\begin{align*}
\Pr\left\{ \log\frac{p_{Y^n|X^n}(Y^n(1)|X^n(2))}{p_{Y^n}(Y^n(1))} > \log M + \delta \right\} \le \frac{\mathrm{e}^{-\delta}}{M}\,.
\end{align*}
\end{Lemma}

\medskip
The following lemma is a modification of the Shannon's bound stated in the previous lemma, and its proof is provided in Appendix~\ref{appendixA+Block}.
\begin{Lemma} \label{lemmaFeinstein}
Suppose we are given a save-and-transmit $(n, M)$-EH code with a length-$m$ saving phase as described in Section~\ref{subsecSaveTransmit}.
Then for each $\gamma \ge 0$, each $\delta>0$ and each~$M\in \mathbb{N}$, we have
\begin{align*}
&\Pr\left\{\left.\left\{ \sum_{k=m+1}^n\log\frac{p_{Y_k|X_k}(\tilde Y_k(1)|X_k(2))}{p_{Y_k}(\tilde Y_k(1))} > \log M + \delta\,\right\} \cap\left\{ |\mathcal{Q}^{(n)}(1)|<\gamma+1\right\}\right| W=1 \right\} \notag\\*
&\quad\le \frac{2\mathrm{e}^{-\delta}}{M}\,\times ((n-m)\sqrt{S+1})^{\gamma+1}. 
\end{align*}
\end{Lemma}
\subsection{A Non-Asymptotic Achievable Rate for Save-and-Transmit}
The following theorem is the main result of this paper. The proof relies on Lemma~\ref{lemmaSavingMismatch} and Lemma~\ref{lemmaFeinstein}, and it will be presented in Section~\ref{sectionThmMainResultBlock}.
\begin{Theorem}\label{thmMainResultContAWGN}
Fix an $\varepsilon\in(0,1)$, fix a natural number $n$, fix a non-negative integer $m < n$, and fix a $\rho\in(0,1)$ such that~\eqref{sufficientlyLargeN} holds. Let $n_m\triangleq n-m$.
Define~$S$, $\alpha$ and~$\beta$ as in~\eqref{defS}, \eqref{defAlpha} and~\eqref{defBeta} respectively.
Let $p_X= \mathcal{N}(x;0, S)$ and let $p_Y=\mathcal{N}(y; 0, S+1)$ be the marginal distribution of~$p_Xq_{Y|X}$, and let $\sigma^2$ and $T$ denote the variance and the third absolute moment of
$\log\frac{q_{Y|X}(Y|X)}{p_{Y}(Y)}$
respectively.
For any $\varepsilon_1>0$ and $\varepsilon_2>0$ such that $\varepsilon_1+\varepsilon_2 =\varepsilon$, if $n$ and~$m$ satisfy
\[
\varepsilon_1 -\frac{T}{\sigma^3\sqrt{n_m}}-\frac{4}{\sqrt{n_m}}>0,
\]
then there exists a save-and-transmit $(n, M)$-EH code with a length-$m$ saving phase which satisfies
\begin{align*}
\log M &\ge \frac{n_m}{2}\log (1+S) + \sqrt{n_m \sigma^2}\,\Phi^{-1}\left(\varepsilon_1 -\frac{T}{\sigma^3\sqrt{n_m}}-\frac{4}{\sqrt{n_m}}\right)  \notag\\*
&\qquad -\left(2S\log 2+\frac{1}{2}\log(1+S) + (8S+1)\log n_m\right)(\gamma(\varepsilon_2)+1) - \log \sqrt{n_m} -1 
\end{align*}
and
\begin{align*}
\Pr\left\{\varphi\big(\tilde Y^n(W)\big)\ne W\right\} \le \varepsilon 
\end{align*}
where
\begin{align*}
\gamma(\varepsilon_2)\triangleq \max\left\{ \frac{\log\frac{1}{\varepsilon_2}}{P\beta+\frac{\alpha^2 \E[E^2]}{2}}-m, 0\right\}.  
\end{align*}
In particular, the probability of seeing more than $\gamma(\varepsilon_2)+1$ mismatch events can be bounded as
\begin{align*}
\Pr\big\{|\mathcal{Q}^{(n)}(W)|\ge \gamma(\varepsilon_2)+1\big\}\le \varepsilon_2. 
\end{align*}
\end{Theorem}

The following corollary is a direct consequence of Theorem~\ref{thmMainResultContAWGN}, and it states a non-asymptotic rate for the save-and-transmit scheme whose second-order term scales as $-O\Big(\frac{1}{\sqrt{n}}\Big)$. The proof of Corollary~\ref{corollaryThmMainResultAWGNsaving} is provided in Appendix~\ref{appendixDblock}.
\begin{Corollary} \label{corollaryThmMainResultAWGNsaving}
Fix an $\varepsilon\in(0,1/2)$, and fix any $\varepsilon_1>0$ and $\varepsilon_2>0$ such that $\varepsilon_1+\varepsilon_2= \varepsilon$. There exists a constant $\kappa>0$ which does not depend on~$n$ such that for all sufficiently large~$n$, we can construct a save-and-transmit $(n, M, \varepsilon)$-EH code which satisfies
\begin{align}
\frac{1}{n}\log M  \ge  \frac{1}{2}\log (1+P) -\sqrt{\frac{(\E[E^2]+3P^2)\log(1+P)\log\frac{1}{\varepsilon_2}}{2nP(P+1)}}   + \sqrt{\frac{P}{(P+1)n}}\,\Phi^{-1}(\varepsilon_1)-\frac{\kappa }{n^{3/4}},\label{st1CorollaryMainResultAWGNsaving}
\end{align}
with $\rho$ being defined as
  \begin{align*}
\rho\triangleq \frac{\sqrt{(P+1)(\E[E^2]+3P^2)\log(1+P)\log\frac{1}{\varepsilon_2}}}{P\sqrt{2nP}} =\Theta\left(\frac{1}{\sqrt{n}}\right),
\end{align*}
the average transmit power~$S$ being defined as in~\eqref{defS},
$\alpha$ and~$\beta$ being defined as in~\eqref{defAlpha} and~\eqref{defBeta} respectively, and the length of saving phase~$m$ being defined as
\begin{align*}
m\triangleq \left\lceil\frac{\log\frac{1}{\varepsilon_2}}{P\beta+\frac{\alpha^2 \E[E^2]}{2}}\right\rceil= \Theta\left(\sqrt{n} \right) 
\end{align*}
In particular, the probability of seeing a mismatch event in the transmission phase can be bounded as
\begin{align*}
\Pr\left\{\bigcup_{k=m+1}^n\left\{\sum_{i=1}^k E_i < \sum_{i=m+1}^k X_i^2  \right\} \right\} \le \varepsilon_2 
\end{align*}
where each term in the union characterizes the event that the accumulated energy collected during the first $k$ time slots is insufficient to
output the desired codeword symbols from time~$m+1$ to time~$k$ during the transmission phase.
\end{Corollary}

\begin{Remark}\label{remarkCorollarySaving}
The parameters~$\rho$ and~$m$ in Corollary~\ref{corollaryThmMainResultAWGNsaving} have been carefully chosen to achieve the second-order scaling~$-O(1/\sqrt{n})$, where the scaling is optimal~\cite[Th.~1]{FTO17}.
Fix any $\varepsilon\in(0, 1/2)$. The best existing lower bound on the second-order term of $\frac{1}{n}\log M$ was derived in~\cite[Th.~1]{FTO17}, which states that there exists a save-and-transmit $(n, M, \varepsilon)$-EH code that satisfies
\begin{align}
\liminf_{n\rightarrow \infty}\frac{1}{\sqrt{n}}\left(\log M -\frac{n}{2}\log(1+P)\right) \ge  -\frac{\log(1+P)}{2P}\sqrt{(\E[E^2]+P^2)\log\frac{1}{\varepsilon_2}}   + \sqrt{\frac{P}{P+1}}\,\Phi^{-1}(\varepsilon_1) \label{existingBest}
\end{align}
for any $\varepsilon_1>0$ and $\varepsilon_2>0$ such that $\varepsilon_1+\varepsilon_2= \varepsilon$. The save-and-transmit scheme investigated in~\cite{FTO17} is similar to that described in Section~\ref{subsecSaveTransmit} except that $S=\E[E]=P$ is assumed in~\cite{FTO17} while $S<\E[E]=P$ is assumed in this work. Note that the second-order term of the best existing lower bound as stated on the right-hand side (RHS) of~\eqref{existingBest} decays as~$-\frac{1}{2}\log(1+P)\sqrt{(1+\frac{\E[E^2]}{P^2})\log\frac{1}{\varepsilon_2}} + \Phi^{-1}(\varepsilon_1)$ as $P$ tends to~$\infty$. On the other hand, it follows from \eqref{st1CorollaryMainResultAWGNsaving} in Corollary~\ref{corollaryThmMainResultAWGNsaving} that the second-order term of our lower bound decays as $-\sqrt{\frac{1}{2}(3+\frac{\E[E^2]}{P^2})\log(1+P)\log\frac{1}{\varepsilon_2}} + \Phi^{-1}(\varepsilon_1)$ as $P$ tends to~$\infty$. Consequently, the second-order term achievable by the save-and-transmit scheme guaranteed by Corollary~\ref{corollaryThmMainResultAWGNsaving} is strictly larger (less negative) than the best existing bound for all sufficiently large~$P>0$. In other words, letting~$S$ be strictly less than instead of equal to~$P$ achieves a higher rate in the high SNR regime.
\end{Remark}
\subsection{A Non-Asymptotic Achievable Rate for Best-Effort} \label{subsecBestEffort}
We call a save-and-transmit scheme a \textit{best-effort scheme} if the length of saving phase equals zero, i.e., $m=0$. By setting~$m=0$, Theorem~\ref{thmMainResultContAWGN} reduces to the following corollary, which states that the best-effort scheme achieves a non-asymptotic rate whose second-order term scales as $-O\Big(\sqrt{\frac{\log n}{n}}\Big)$. The proof of Corollary~\ref{corollaryThmMainResultAWGN} is provided in Appendix~\ref{appendixBblock}.
\begin{Corollary} \label{corollaryThmMainResultAWGN}
Fix an $\varepsilon\in(0,1/2)$, and fix any $\varepsilon_1>0$ and $\varepsilon_2>0$ such that $\varepsilon_1+\varepsilon_2= \varepsilon$.
Define
\begin{align}
\lambda_1\triangleq 2P\log 2+\frac{1}{2}\log(1+P) \label{defLambda1}
\end{align}
and
\begin{align}
\lambda_2\triangleq 8P+1. \label{defLambda2}
\end{align}
 There exists a constant $\kappa>0$ which does not depend on~$n$ such that for all sufficiently large~$n$, we can construct a best-effort $(n, M, \varepsilon)$-EH code
with
 \begin{align*}
\rho\triangleq \frac{\sqrt{(\lambda_1 + \lambda_2\log n)(P+1)(\E[E^2]+3P^2)\log\frac{1}{\varepsilon_2}}}{P^{3/2}}\times\frac{1}{\sqrt{n}}=\Theta\left(\sqrt{\frac{\log n}{n}} \right) 
\end{align*}
and
\begin{align*}
S=P(1-\rho)=P-\Theta\left(\sqrt{\frac{\log n}{n}} \right)
\end{align*}
 which satisfies
\begin{align}
\frac{1}{n}\log M  \ge  \frac{1}{2}\log (1+P) -\sqrt{\frac{(\lambda_1 + \lambda_2\log n)(\E[E^2]+3P^2)\log\frac{1}{\varepsilon_2}}{P(P+1)}}\times\frac{1}{\sqrt{n}}    - \sqrt{\frac{P}{(P+1)n}}\,\Phi^{-1}(\varepsilon_1)-\frac{\kappa \log n}{n}.\label{st1CorollaryMainResultAWGN}
\end{align}
In particular, the probability of seeing more than
\begin{align*}
\gamma(\varepsilon_2)\triangleq \frac{\log\frac{1}{\varepsilon_2}}{P\beta+\frac{\alpha^2 \E[E^2]}{2}} =\Theta\Big(\sqrt{\frac{n}{\log n}}\,\Big)
\end{align*}
 mismatch events can be bounded as
\begin{align*}
\Pr\big\{|\mathcal{Q}^{(n)}(W)|\ge \gamma(\varepsilon_2)+1\big\}\le \varepsilon_2. 
\end{align*}
\end{Corollary}
\begin{Remark} \label{remarkBestEffort}
Although the achievable second-order scaling for best-effort in Corollary~\ref{corollaryThmMainResultAWGN} is not optimal (the optimal scaling is~$-O(1/\sqrt{n})$~\cite[Th.~1]{FTO17}), it is a significant improvement compared to the state of the art~\cite[Sec.~V]{ozel12} where the achievable second-order scaling therein for best-effort is $-o(1)$.
\end{Remark}

\section{The Block Energy Arrival Model} \label{sectionBlockEH}
In this section, we generalize our achievable rates for save-and-transmit and best-effort to the block energy arrival model~\cite{ZhangLau14,shavivOzgur18,FTO17}, which is useful for modeling practical scenarios when the energy-arrival process (e.g., solar energy, wind energy, ambient radio-frequency (RF) energy, etc.) evolves at a slower timescale compared to the transmission process.
\subsection{Block Energy Arrivals}
We follow the formulation in~\cite{FTO17}, which assumes that $\{E_i\}_{i=1}^\infty$ arrive at the buffer in a block-by-block manner as follows: For each $\ell\in\mathbb{N}$, let
 \begin{equation}
 b_\ell \triangleq (\ell-1)L \label{blockIndex}
 \end{equation}
such that $b_\ell+1$ is the index of the first channel use within the $\ell^{\text{th}}$ block of energy arrivals, where~$L$ denotes the length of each block. The EH random variables that mark the starting positions of the blocks (i.e., $\{E_{b_\ell+1}\}_{\ell=1}^\infty$) are assumed to be i.i.d.\ random variables where $E_1=E$ satisfies~\eqref{defEnergyP}. In addition, we assume
 \begin{equation*}
E_{b_\ell+1}=E_{b_\ell+2}=\ldots = E_{b_{\ell}+L} 
 \end{equation*}
 for all $\ell\in\mathbb{N}$. In other words, the harvested energy in each channel use within a block remains constant while the harvested energy across different blocks is characterized by a sequence of i.i.d.\ random variables with mean equal to~$P$.
By construction, we have the following for each $k\in\{1, 2, \ldots, n\}$ and all $e^k\in\mathbb{R}_+^k$,
\begin{align*}
p_{E_k|E^{k-1}}(e_k| e^{k-1})=
\begin{cases}
p_{E_1}(e_k) & \text{if $k=b_\ell+1$ for some $\ell\in\mathbb{N}$,}\\
\mathbf{1}\{e_k=e_{k-1}\} & \text{otherwise.}
\end{cases}
\end{align*}
The length of each energy-arrival block~$L$ is assumed to remain constant or grow sublinearly in~$n$.
%
\subsection{Blockwise Save-and-Transmit} \label{subsecSaveTransmitBlock}
Fix a blocklength~$n$ and choose an~$L = o(n)$. Choose a positive real number $S<P=\E[E]$ and let $p_X$ be as defined in~\eqref{defDistPX} such that $S= \E_{p_X}[X^2]$.
  The codebook consists of~$M$ mutually independent random codewords denoted by $\{X^n(w)\,|\, w\in\mathcal{W}\}$, which are constructed as described in Section~\ref{subsecSaveTransmit}. Suppose $W=w$ and $E^n=e^n$. Then, the transmitter uses the following \emph{blockwise save-and-transmit $(n, M)$-EH code} with encoding functions~$\{f_k\}_{k=1}^{\bar n}$ and decoding function~$\varphi$ where $\bar n\triangleq\lceil n/L\rceil$. The saving phase consists of~$m$ blocks of~$L$ consecutive time slots. Define $f_1, f_2, \ldots, f_{\bar n}$ in a recursive manner where
  \begin{align}
f_\ell(w, e^{b_\ell+1}) \triangleq  \begin{cases}
\left( X_{b_{\ell}+1}(w), X_{b_{\ell}+2}(w), \ldots, X_{b_{\ell}+L}(w) \right)& \parbox[t]{4.5 in}{if $m< \ell<\bar n$ and\\ \text{ }$\sum\limits_{j=1}^{L}\big(X_{b_{\ell}+j}(w)\big)^2 \le \sum\limits_{k=1}^{b_{\ell}+1} e_k - \sum\limits_{i=1}^{\ell-1} \big\|f_i(w, e^{b_i+1})\big\|^2 $,}\\
\left( X_{b_{\bar n}+1}(w), X_{b_{\bar n}+2}(w), \ldots, X_{n}(w) \right)& \parbox[t]{4.5 in}{if $\ell= \bar n$ and\\ \text{ }$\sum\limits_{k=b_{\bar n}+1}^n\big(X_k(w)\big)^2 \le \sum\limits_{k=1}^{b_{\bar n}+1} e_k - \sum\limits_{i=1}^{\bar n-1} \big\|f_i(w, e^{b_i+1})\big\|^2 $,}
\\
 \underbrace{(0, 0, \ldots, 0)}_{\min\{L, n-b_{\ell}\} \text{ times}} & \text{otherwise.}
\end{cases} \label{defSavingEncodingBlock}
\end{align}
In other words, the transmitter outputs a block of~$L$ symbols $\left( X_{b_{\ell}+1}(w), X_{b_{\ell}+2}(w), \ldots, X_{b_{\ell}+L}(w) \right)$ in the transmission phase during time $b_{\ell}+1$ to $b_{\ell}+L$ if the energy in the battery at time~$b_{\ell}+1$ (i.e., $\sum_{k=1}^{b_{\ell}+1} e_{k} - \sum_{i=1}^{\ell-1} \big\|f_i(w, e^{b_i+1})\big\|^2$) can support the transmission of the whole block of symbols starting at time $b_\ell + 1$. If $L=1$, the blockwise save-and-transmit scheme defined by~\eqref{defSavingEncodingBlock} reduces to the save-and-transmit scheme presented in Section~\ref{subsecSaveTransmit} defined by~\eqref{defSaveEncoding}. Let $\tilde X_k(W)$ be the symbol transmitted at time~$k$ for each $k\in\{1, 2, \ldots, n\}$ such that
\begin{equation*}
(\tilde X_{b_{\ell}+1}(W), \tilde X_{b_{\ell} +2}(W), \ldots, \tilde X_{\min\{b_{\ell}+L, n\}}(W) )\triangleq f_\ell(W, E^{b_\ell+1}) 
\end{equation*}
for each $\ell\in\{1, 2, \ldots, \bar n\}$.
Upon receiving $\tilde Y^n (W)\triangleq (\tilde Y_1(W), \tilde Y_2(W), \ldots, \tilde Y_n(W))$ where $\tilde Y_k(W)$ is generated according to~\eqref{channelLawDeltaAWGN},
the receiver declares that $\varphi(\tilde Y^n(W))=j$ if $j$ is the unique integer in~$\mathcal{W}$ that satisfies
\begin{align}
\sum_{k=mL+1}^n \log\frac{q_{Y|X}(\tilde Y_k(W)|X_k(j))}{p_{Y}(\tilde Y_k(W))} \ge \log \xi, \label{defSaveDecodingAWGNblock}
\end{align}
where $p_Y$ is the marginal distribution of $p_X q_{Y|X}$ and $\log \xi$ is an arbitrary threshold to be carefully chosen later (cf.\ \eqref{defXiAWGNblock}).
Otherwise, the receiver chooses $\varphi(\tilde Y^n(W))\in\mathcal{W}$ according to the uniform distribution. The decoding is successful if $j=W$.

The following lemma is an extension of Lemma~\ref{lemmaSavingMismatch} which states an upper bound on the probability of seeing more than $L\gamma+1$ mismatched positions in the transmission phase. The proof of Lemma~\ref{lemmaMismatchBlock} is contained in Appendix~\ref{appendixAblock}.
\begin{Lemma}\label{lemmaMismatchBlock}
Fix any~$n$, any~$L\le n$ and any $\rho\in(0, 1)$ such that~\eqref{sufficientlyLargeN} holds,
 and fix a blockwise save-and-transmit $(n, M)$-EH code with~$S$ being defined as in~\eqref{defS}.
Define
  \begin{equation}
\alpha\triangleq \frac{2\rho  P}{L\E[E^2]+3S^2} \label{defAlphaBlock}
 \end{equation}
 and
  \begin{equation}
\beta\triangleq \frac{\alpha}{1+63\alpha S}. \label{defBetaBlock}
 \end{equation}
For any~$\gamma\in\mathbb{R}_+$, we have
\begin{align}
\Pr\left\{|\mathcal{Q}^{(n)}(W)| \ge L\gamma+1\right\} \le  \mathrm{e}^{-L(m+\gamma)\big(P\beta+\frac{L\alpha^2 \E[E^2]}{2}\big)}. \label{stLemmaMismatchBlock}
\end{align}
\end{Lemma}
\begin{Remark}
In the proof of Lemma~\ref{lemmaMismatchBlock} in Appendix~\ref{appendixAblock}, an important step is analyzing the escape probability~\eqref{appendixLemma1Eq0-Block} of the Markov process $\left\{\sum_{i=1}^{m} L E_{b_i+1}+\sum_{i=m+1}^{m+k} \big( L E_{b_i+1}- \sum_{\ell=1}^L X_{b_i+\ell}^2\big)\right\}_{k=1}^\tau$ where $\tau$ is the stopping time when the value of the Markov process hits any negative number $a<0$.
\end{Remark}
\medskip

The following lemma is a generalization of Lemma~\ref{lemmaFeinstein}. The proof of Lemma~\ref{lemmaFeinsteinBlock} is contained in Appendix~\ref{appendixA+Block}. 
\begin{Lemma} \label{lemmaFeinsteinBlock}
Suppose we are given a blockwise save-and-transmit $(n, M)$-EH code with a saving phase of length~$mL$ as described in Section~\ref{subsecSaveTransmitBlock}.
Then for each natural number $L<n/m$, each $\gamma \ge 0$, each $\delta>0$ and each~$M\in \mathbb{N}$, we have
\begin{align}
&\Pr\left\{\left.\left\{ \sum_{k=mL+1}^n\log\frac{p_{Y_k|X_k}(\tilde Y_k(1)|X_k(2))}{p_{Y_k}(\tilde Y_k(1))} > \log M + \delta\,\right\} \cap\left\{ |\mathcal{Q}^{(n)}(1)|<L\gamma+1 \right\}\right| W=1 \right\} \notag\\*
&\qquad \le \frac{2\mathrm{e}^{-\delta}}{M}\,\times \big((n-mL)(S+1)^{L/2}\big)^{\gamma+1}. \label{stLemmaFeinsteinBlock}
\end{align}
\end{Lemma}

\subsection{A Non-Asymptotic Achievable Rate for Blockwise Save-and-Transmit}
The following theorem is the main result under the block energy arrival model. The proof relies on Lemma~\ref{lemmaMismatchBlock} and Lemma~\ref{lemmaFeinstein*}, and it will be provided in Section~\ref{sectionThmMainResultBlock}.
\begin{Theorem}\label{thmMainResultContAWGNsavingBlock}
Fix an $\varepsilon\in(0,1)$, fix a natural number $n\ge 2$, fix a natural number $L\le n$, fix a non-negative integer $m < n$, and fix a $\rho\in(0,1)$ such that~\eqref{sufficientlyLargeN} holds. Let $n_m\triangleq n-mL$.
Define $S$, $\alpha$ and $\beta$ as in~\eqref{defS}, \eqref{defAlphaBlock} and~\eqref{defBetaBlock} respectively.
Let $p_X= \mathcal{N}(x;0, S)$ and let $p_Y=\mathcal{N}(y; 0, S+1)$ be the marginal distribution of~$p_Xq_{Y|X}$, and let $\sigma^2$ and $T$ denote the variance and the third absolute moment of
$ \log\frac{q_{Y|X}(Y|X)}{p_{Y}(Y)}$
respectively.
For any $\varepsilon_1>0$ and $\varepsilon_2>0$ such that $\varepsilon_1+\varepsilon_2 =\varepsilon$, if $n$ and~$m$ satisfy
\begin{align}
\varepsilon_1 -\frac{T}{\sigma^3\sqrt{n_m}}-\frac{4}{\sqrt{n_m}}>0, \label{assumptionThm}
\end{align}
 then there exists a blockwise save-and-transmit $(n, M)$-EH code with a saving phase of length-$mL$ such that
\begin{align}
\log M &\ge \frac{n_m}{2}\log (1+S) + \sqrt{n_m \sigma^2}\,\Phi^{-1}\left(\varepsilon_1 -\frac{T}{\sigma^3\sqrt{n_m}}-\frac{4}{\sqrt{n_m}}\right) \notag\\*
&\qquad -\left(L\Big(2S\log 2+\frac{1}{2}\log(1+S)\Big) + (8S+1)\log n_m\right) (\gamma(\varepsilon_2)+1) - \log \sqrt{n_m} -1 \label{st1ThmMainResultAWGNblock}
\end{align}
and
\begin{align}
\Pr\left\{\varphi\big(\tilde Y^n(W)\big)\ne W\right\} \le \varepsilon \label{st2ThmMainResultAWGNblock}
\end{align}
where
\begin{align}
\gamma(\varepsilon_2)\triangleq \max\left\{\frac{\log\frac{1}{\varepsilon_2}}{LP\beta+\frac{L^2\alpha^2 \E[E^2]}{2}}-m, 0\right\}. \label{defGammaBlock}
\end{align}
In particular, the probability of seeing more than $L\gamma(\varepsilon_2)+1$ mismatch events can be bounded as
\begin{align}
\Pr\big\{|\mathcal{Q}^{(n)}(W)|\ge L\gamma(\varepsilon_2)+1\big\}\le \varepsilon_2. \label{st3ThmMainResultAWGNblock}
\end{align}
\end{Theorem}
The following corollary is a direct consequence of Theorem~\ref{thmMainResultContAWGNsavingBlock}, and it states a non-asymptotic rate for the blockwise save-and-transmit scheme whose second-order term scales as $-O\Big(\sqrt{\frac{L}{n}}\Big)$. The proof of Corollary~\ref{corollaryThmMainResultAWGNsavingBlock} is provided in Appendix~\ref{appendixDblock}.
\begin{Corollary} \label{corollaryThmMainResultAWGNsavingBlock}
Fix an $\varepsilon\in(0,1/2)$, and fix any $\varepsilon_1>0$ and $\varepsilon_2>0$ such that $\varepsilon_1+\varepsilon_2= \varepsilon$. Suppose $L=o(n)$. There exists a constant $\kappa>0$ which does not depend on~$n$ such that for all sufficiently large~$n$, we can construct a blockwise save-and-transmit $(n, M, \varepsilon)$-EH code that satisfies
\begin{align}
\frac{1}{n}\log M  \ge  \frac{1}{2}\log (1+P) -\sqrt{\frac{(L\E[E^2]+3P^2)\log(1+P)\log\frac{1}{\varepsilon_2}}{2nP(P+1)}}   + \sqrt{\frac{P}{(P+1)n}}\,\Phi^{-1}(\varepsilon_1)-\kappa\max\left\{\frac{ L^{1/4}}{n^{3/4}}, \frac{L}{n}\right\},\label{st1CorollaryMainResultAWGNsavingBlock}
\end{align}
 with $\rho$ being defined as
  \begin{align}
\rho \triangleq \frac{\sqrt{(P+1)(L\E[E^2]+3P^2)\log(1+P)\log\frac{1}{\varepsilon_2}}}{P\sqrt{2nP}} =\Theta\left(\sqrt{\frac{L}{n}}\right), \label{defRhoSavingBlock}
\end{align}
the average transmit power~$S$ being defined as in~\eqref{defS}, $\alpha$ and~$\beta$ being defined as in~\eqref{defAlphaBlock} and~\eqref{defBetaBlock} respectively, and the length of saving phase~$mL$ being defined as
\begin{align}
m L\triangleq L\left\lceil\frac{\log\frac{1}{\varepsilon_2}}{LP\beta+\frac{L^2\alpha^2 \E[E^2]}{2}}\right\rceil= \Theta(\sqrt{nL}). \label{defmSavingBlock}
\end{align}
In particular, the probability of seeing a mismatch event in the transmission phase can be bounded as
\begin{align}
\Pr\left\{\bigcup_{k=mL+1}^n\left\{\sum_{i=1}^k E_i < \sum_{i=mL+1}^k X_i^2  \right\} \right\} \le \varepsilon_2. \label{st2CorollaryMainResultAWGNsavingBlock}
\end{align}
\end{Corollary}
\medskip
The following result is a direct consequence of Corollary~\ref{corollaryThmMainResultAWGNsavingBlock}.
\begin{Theorem} \label{thmBlockSave}
Fix any $\varepsilon\in(0,1/2)$. Suppose $L=\omega(1)\cap o(n)$, i.e., $\lim_{n\rightarrow\infty}\frac{1}{L}=\lim_{n\rightarrow\infty}\frac{L}{n}=0$. Then for all sufficiently large~$n$,  there exists a blockwise save-and-transmit $(n, M, \varepsilon)$-EH code such that
\begin{align}
\frac{1}{n}\log M &\ge  \frac{1}{2}\log (1+P) -\sqrt{\frac{\E[E^2]\log(1+P)\log\frac{1}{\varepsilon}}{2P(P+1)}}\times \sqrt{\frac{L}{n}}  - o\left(\sqrt{\frac{L}{n}}\right).\label{st1ThmBlockSave}
\end{align}
\end{Theorem}
\begin{IEEEproof}
It follows from Corollary~\ref{corollaryThmMainResultAWGN} that for all sufficiently large~$n$, there exists a blockwise save-and-transmit $(n, M, \varepsilon)$-EH code that satisfies~\eqref{st1CorollaryMainResultAWGNsavingBlock}, which together with hypothesis regarding~$L$ implies~\eqref{st1ThmBlockSave}.
\end{IEEEproof}
\medskip
\begin{Remark}\label{remarkCorollarySavingBlock}
Fix any $\varepsilon \in (0,1/2)$ and fix any $L=\omega(1)\cap o(n)$. The best existing lower bound on the second-order term of $\frac{1}{n}\log M$ was derived in~\cite[Th.~1]{FTO17}, which states that there exists a save-and-transmit $(n, M, \varepsilon)$-EH code that satisfies
\begin{align}
\liminf_{n\rightarrow \infty}\frac{1}{\sqrt{Ln}}\left(\log M -\frac{n}{2}\log(1+P)\right) \ge  -\frac{\log(1+P)}{2P}\sqrt{(\E[E^2]+P^2)\log\frac{1}{\varepsilon}}. \label{existingBestBlock}
\end{align}
The blockwise save-and-transmit scheme investigated in~\cite{FTO17} is similar to that described in Section~\ref{subsecSaveTransmitBlock} except that $S=\E[E]=P$ is assumed in~\cite{FTO17} while $S<\E[E]=P$ is assumed in this work.
Note that the second-order term of the best existing lower bound as stated on the RHS of~\eqref{existingBestBlock} decays as\linebreak $-\frac{1}{2}\log(1+P)\sqrt{(1+\frac{\E[E^2]}{P^2})\log\frac{1}{\varepsilon}}$ as $P$ tends to~$\infty$. On the other hand, it follows from \eqref{st1ThmBlockSave} in Theorem~\ref{thmBlockSave} that the second-order term of our lower bound decays as $-\sqrt{\frac{\E[E^2]}{2P^2}\log(1+P)\log\frac{1}{\varepsilon}}$ as $P$ tends to~$\infty$. Consequently, the second-order term achievable by the save-and-transmit scheme guaranteed by Theorem~\ref{thmBlockSave} is strictly larger (less negative) than the best existing bound for all sufficiently large~$P>0$.
\end{Remark}
\subsection{A Non-Asymptotic Achievable Rate for Blockwise Best-Effort} \label{subsecBestEffortBlock}
We call a blockwise save-and-transmit scheme a \textit{blockwise best-effort scheme} if the length of saving phase equals zero, i.e., $m=0$. By setting~$m=0$, Theorem~\ref{thmMainResultContAWGNsavingBlock} reduces to the following corollary, which states that blockwise best-effort achieves a non-asymptotic rate whose second-order term scales as $-O\Big(\sqrt{\frac{\max\{\log n, L\}}{n}}\Big)$. The proof of Corollary~\ref{corollaryThmMainResultAWGNblock} is provided in Appendix~\ref{appendixBblock}.
\begin{Corollary} \label{corollaryThmMainResultAWGNblock}
Fix an $\varepsilon\in(0,1/2)$, and fix any $\varepsilon_1>0$ and $\varepsilon_2>0$ such that $\varepsilon_1+\varepsilon_2= \varepsilon$. Define~$\lambda_1$ and~$\lambda_2$ as in~\eqref{defLambda1} and~\eqref{defLambda2} respectively.
 There exists a constant $\kappa>0$ which does not depend on~$n$ such that for all sufficiently large~$n$ and any $L\le n$, we can construct a blockwise best-effort $(n, M, \varepsilon)$-EH code
with
 \begin{align}
\rho &\triangleq \frac{\sqrt{(\lambda_1 L + \lambda_2\log n)(P+1)(L\E[E^2]+3P^2)\log\frac{1}{\varepsilon_2}}}{P\sqrt{PLn}} \notag\\*
& = \Theta\left(\sqrt{\frac{\max\{\log n,L\}}{n}} \right)\label{defRhoBlock}
\end{align}
and
\begin{align*}
S=P(1-\rho )=P-\Theta\left(\sqrt{\frac{\max\{\log n,L\}}{n}} \right)
\end{align*}
 which satisfies
\begin{align}
\frac{1}{n}\log M &\ge  \frac{1}{2}\log (1+P) -\sqrt{\frac{(\lambda_1 L + \lambda_2 \log n)(L\E[E^2]+3P^2)\log\frac{1}{\varepsilon_2}}{LP(P+1)n}}   \notag\\*
&\qquad- \sqrt{\frac{P}{(P+1)n}}\,\Phi^{-1}(\varepsilon_1)-\frac{\kappa \max\{\log n,L\}}{n}.\label{st1CorollaryMainResultAWGNblock}
\end{align}
In particular, the probability of seeing more than
\begin{align}
\gamma(\varepsilon_2)\triangleq \frac{\log\frac{1}{\varepsilon_2}}{LP\beta+\frac{L^2\alpha^2 \E[E^2]}{2}} =\Theta\Big(\sqrt{\frac{n}{\max\{\log n,L\}}}\,\Big) \label{defGammaBestEffortBlock}
\end{align}
 mismatch events can be bounded as
$
\Pr\big\{|\mathcal{Q}^{(n)}(W)|\ge \gamma(\varepsilon_2)+1\big\}\le \varepsilon_2$. 
\end{Corollary}
\medskip
\begin{Remark} \label{remark6}
The parameters~$\rho$ and $\gamma(\varepsilon_2)$ in Corollary~\ref{corollaryThmMainResultAWGNblock} have been optimized to achieve the second-order scaling $-O\Big(\sqrt{\frac{\max\{\log n, L\}}{n}}\Big)$.
\end{Remark}
\medskip
The following result is a direct consequence of Corollary~\ref{corollaryThmMainResultAWGNblock}.
\begin{Theorem} \label{thmBlockBestEffort}
Fix any $\varepsilon\in(0,1/2)$. Suppose $L=\omega(\log n)\cap o(n)$, i.e., $\lim_{n\rightarrow\infty}\frac{\log n}{L}=\lim_{n\rightarrow\infty}\frac{L}{n}=0$. Then for all sufficiently large~$n$,  there exists a blockwise best-effort $(n, M, \varepsilon)$-EH code such that
\begin{align}
\frac{1}{n}\log M &\ge  \frac{1}{2}\log (1+P) -\sqrt{\frac{\big(2P\log 2+\frac{1}{2}\log(1+P)\big)\E[E^2]\log\frac{1}{\varepsilon}}{P(P+1)}}\times \sqrt{\frac{L}{n}}  - o\left(\sqrt{\frac{L}{n}}\right).\label{st1ThmBlockBestEffort}
\end{align}
\end{Theorem}
\begin{IEEEproof}
It follows from Corollary~\ref{corollaryThmMainResultAWGNblock} that for all sufficiently large~$n$, there exists a blockwise best-effort $(n, M, \varepsilon)$-EH code that satisfies~\eqref{st1CorollaryMainResultAWGNblock} where $\varepsilon_1$ and $\varepsilon_2$ are chosen to be~$\varepsilon/n$ and~$\varepsilon(1-1/n)$ respectively, which together with the definitions of~$\lambda_1$ and~$\lambda_2$ in~\eqref{defLambda1} and~\eqref{defLambda2} and the hypothesis regarding~$L$ implies~\eqref{st1ThmBlockBestEffort}.
\end{IEEEproof}
\medskip
\begin{Remark}  \label{remarkBestEffortBlock}
If $L=\omega(\log n)\cap o(n)$, the achievable second-order scaling for blockwise best-effort in Theorem~\ref{thmBlockBestEffort} is $O\big(\sqrt{\frac{L}{n}}\big)$ which is optimal~\cite[Th.~1]{FTO17}). However, we can see from Theorem~\ref{thmBlockSave} and Theorem~\ref{thmBlockBestEffort} that blockwise best-effort always achieves a smaller (more negative) coefficient for the second-order term than save-and-transmit.
\end{Remark}

\section{Proofs of Theorem~\ref{thmMainResultContAWGN} and Theorem~\ref{thmMainResultContAWGNsavingBlock}}\label{sectionThmMainResultBlock}
Since save-and-transmit defined in Section~\ref{subsecSaveTransmit} is a special case of blockwise save-and-transmit defined in Section~\ref{subsecSaveTransmitBlock} with $L=1$ and Theorem~\ref{thmMainResultContAWGN} is a special case of Theorem~\ref{thmMainResultContAWGNsavingBlock} with $L=1$, it suffices to prove Theorem~\ref{thmMainResultContAWGNsavingBlock}.

Fix an $\varepsilon\in(0,1)$ and any $\varepsilon_1>0$ and $\varepsilon_2>0$ such that $\varepsilon_1+\varepsilon_2= \varepsilon$. Fix an $n\in\mathbb{N}$, an $L<n$ and a $\rho\in(0,1)$ that satisfies~\eqref{sufficientlyLargeN}. Consider a blockwise save-and-transmit $(n,M)$-code described in Section~\ref{subsecSaveTransmitBlock} where the corresponding $S$ and $p_X$ are defined according to~\eqref{defS} and~\eqref{defDistPX} respectively. In addition, let $p_Y(y)=\mathcal{N}(y; 0, S+1)$ be the marginal distribution of~$p_Xq_{Y|X}$,
 and define $\alpha$, $\beta$ and $\gamma(\varepsilon_2)$ as in \eqref{defAlphaBlock}, \eqref{defBetaBlock} and \eqref{defGammaBlock} respectively.
 Consider the probability of decoding error
 \begin{align}
 &\Pr\left\{\big\{\varphi\big(\tilde Y^n(W)\big)\ne W\big\}\right\} \notag\\*
 &\quad\le \Pr\left\{\big\{\varphi\big(\tilde Y^n(W)\big)\ne W\big\}\cap \big\{|\mathcal{Q}^{(n)}(W)|< L\gamma(\varepsilon_2)+1\big\} \right\} + \varepsilon_2 \label{MainResultContProofEq2AWGNblock}
 \end{align}
which is due to the union bound and the following fact by Lemma~\ref{lemmaMismatchBlock} (Lemma~\ref{lemmaSavingMismatch} suffices for the case~$L=1$) and the definition of $\gamma(\varepsilon_2)$ in~\eqref{defGammaBlock}:
\begin{align}
\Pr\big\{|\mathcal{Q}^{(n)}(W)|\ge L\gamma(\varepsilon_2)+1\big\}\le \mathrm{e}^{-\log\frac{1}{\varepsilon_2}}= \varepsilon_2.  \label{MainResultContProofEq2AWGNstepAblock}
\end{align}
Recall that $n_m=n-mL$ and $b_\ell+1$ (which was defined in~\eqref{blockIndex}) denotes the first channel use within the $\ell^{\text{th}}$ block of energy arrivals.
Using the convention that $X_{k}(1)=0$ deterministically for all $k>n$, it follows from the code construction that
 \begin{align}
& \Pr\left\{\max_{\ell\in\{m+1, m+2, \ldots, \lceil n/L\rceil\}}\big\|(X_{b_\ell+1}(1), X_{b_\ell+2}(1), \ldots, X_{b_\ell+L}(1))\big\|^2 \ge 2S(L\log 2 + 3\log n_m)\right\} \notag\\*
& \quad \le \frac{n_m}{L} \,\Pr_{p_{X^n}}\left\{ \sum_{k=1}^L X_k^2\ge 2S(L\log 2 + 3\log n_m)\right\} \notag\\
&\quad =  \frac{n_m}{L} \, \Pr_{p_{X^n}}\left\{\mathrm{e}^{ \sum_{k=1}^L\frac{ X_k^2}{4S}}\ge  n_m^{3/2} 2^{L/2}\right\}\notag\\
&\quad \le \frac{n_m}{L}\times \frac{1}{n_m^{3/2} 2^{L/2}}\,\left(\E_{p_X}\left[\mathrm{e}^{\frac{X^2}{4S}}\right]\right)^L  \label{MainResultContProofEq3AWGNstepAblock}\\*
&\quad \le \frac{1}{\sqrt{n_m}},\label{MainResultContProofEq3AWGNblock}
 \end{align}
 where~\eqref{MainResultContProofEq3AWGNstepAblock} follows from Markov's inequality and~\eqref{MainResultContProofEq3AWGNblock} is due to the fact that $X\sim\mathcal{N}(x; 0, S)$.
To simplify notation, define
\begin{align}
\Delta\triangleq L\log2 + 3\log n_m  \label{defDelta}
\end{align}
and
\begin{align}
\tilde{\Delta}\triangleq L\log2 + 4\log n_m, \label{defDelta+}
\end{align}
and define the events
\begin{align}
\mathcal{E}_1 \triangleq\big\{|\mathcal{Q}^{(n)}(1)|< L\gamma(\varepsilon_2) +1 \big\} \label{defE1block}
\end{align}
and
\begin{align*}
\mathcal{E}_2 \triangleq\left\{\max_{\ell\in\{m+1, m+2, \ldots, \lceil n/L\rceil\}}\sum_{j=1}^L(X_{b_\ell+j}(1))^2 < 2S\Delta\right\}. 
\end{align*}
In addition, define
\begin{align}
\imath(a;b)\triangleq\log\frac{q_{Y|X}(b|a)}{p_{Y}(b)}= \frac{1}{2}\log (1+S) +\frac{-S^2 (b-a)^2 + 2a(b-a)+a^2}{2(S+1)} \label{infoSpectrumAWGN}
\end{align}
for all $(a,b)\in\mathbb{R}_+^2$
where $\imath(a;b)$ is used in the decoding rule specified by~\eqref{defSaveDecodingAWGNblock}.
Following~\eqref{MainResultContProofEq2AWGNblock} and
letting $\xi>0$ be an arbitrary positive number to be determined later in~\eqref{defXiAWGNblock}, we obtain from the symmetry of the codebook, the encoding rule~\eqref{defSavingEncodingBlock}, the decoding rule~\eqref{defSaveDecodingAWGNblock}, the union bound and~\eqref{MainResultContProofEq3AWGNblock} that
\begin{align}
&\Pr\left\{\big\{\varphi\big(\tilde Y^n(W)\big)\ne W\big\}\cap \{|\mathcal{Q}^{(n)}(W)|< L\gamma(\varepsilon_2)+1\} \right\}\notag\\*
&\quad = \Pr\left\{\left.\big\{\varphi\big(\tilde Y^n(1)\big)\ne 1\big\}\cap \{|\mathcal{Q}^{(n)}(1)|< L\gamma(\varepsilon_2)+1\} \right|W=1\right\} \notag\\
& \quad \le \Pr\left\{\left.\left(\bigg\{\sum_{k=mL+1}^n \imath(X_k(1); \tilde Y_k(1))  < \log \xi\bigg\}\cup\bigcup_{i=2}^M \left\{\sum_{k=mL+1}^n \imath(X_k(i); \tilde Y_k(1))  \ge \log \xi\right\}\right)\cap \mathcal{E}_1 \right| W=1\right\} \notag\\
& \quad \le \Pr\left\{\left.\bigg\{\sum_{k=mL+1}^n \imath(X_k(1); \tilde Y_k(1))  < \log \xi\bigg\}\cap \mathcal{E}_1 \cap \mathcal{E}_2 \right|W=1  \right\}\notag \\*
 & \quad
 \qquad+  \Pr\left\{\left.\bigcup_{i=2}^M\bigg\{\sum_{k=mL+1}^n \imath(X_k(i); \tilde Y_k(1)) \ge \log \xi\bigg\} \cap \mathcal{E}_1\cap \mathcal{E}_2\right|W=1  \right\} +  \frac{1}{\sqrt{n_m}}\,. \label{MainResultContProofEq5AWGNblock}
\end{align}
 In order to bound the first term in~\eqref{MainResultContProofEq5AWGNblock}, we consider
\begin{align}
&\Pr\left\{\left.\bigg\{\sum_{k=mL+1}^n \imath(X_k(1); \tilde Y_k(1))  < \log \xi\bigg\}\cap \mathcal{E}_1 \cap \mathcal{E}_2 \right|W=1 \right\} \notag\\*
& = \Pr\left\{\left.\bigg\{\sum_{k=mL+1}^n \!\!\!\!\imath(X_k(1);  Y_k(1)) + \sum_{k\in\mathcal{Q}^{(n)}(1)}\!\!\!\!\left(\imath(X_k(1); \tilde Y_k(1))-\imath(X_k(1);  Y_k(1))\right) < \log \xi\bigg\}\cap \mathcal{E}_1 \cap \mathcal{E}_2\right|W=1  \right\}.  \label{MainResultContProofEq6AWGNblock}
\end{align}
Recall from~\eqref{defChannelLawIntro} that $p_{Z^n}(z^n)= \prod_{k=1}^n \mathcal{N}(z_k;0, 1)$. Conditioned on the event $\{\mathcal{Q}^{(n)}(1)=\mathcal{A}\}$, we consider the chain of inequalities below for each block of~$L$ consecutive mismatched positions in $\mathcal{A}$ denoted by $b_\ell+1, b_\ell+2, \ldots, b_\ell+L$:
\begin{align}
&\Pr\left\{\left.\Bigg\{\sum_{k=b_\ell+1}^{b_\ell+L} \imath(X_k(1); \tilde Y_k(1))-\imath(X_k(1);  Y_k(1)) \le - 2S\tilde{\Delta}\Bigg\}\cap \mathcal{E}_2\right|W=1, \mathcal{Q}^{(n)}(1)=\mathcal{A} \right\}\notag\\*
&\quad = \Pr_{ p_{Z^n}p_{X^n|W=1, \mathcal{Q}^{(n)}=\mathcal{A}}}\left\{\bigg\{\sum_{k=b_\ell+1}^{b_\ell+L} \frac{2SX_kZ_k-(S+2)X_k^2}{2(S+1)} \le  - 2S\tilde{\Delta}\bigg\} \cap \left\{\sum_{j=1}^L X_{b_\ell+j}^2< 2S\Delta\right\} \right\} \label{MainResultContProofEq7AWGNstepAblock}\\
&\quad \le \sup_{x^L\,:\, \|x^L\|^2< 2S\Delta} \Pr_{p_{Z^n}}\left\{\sum_{k=1}^L\frac{2Sx_kZ_k-(S+2)x_k^2}{2(S+1)} \le - 2S\tilde{\Delta}\right\} \notag\\
&\quad = \sup_{x^L\,:\, \|x^L\|^2< 2S\Delta} \Pr_{p_Z^n}\left\{\mathrm{e}^{-\sum_{k=1}^L x_kZ_k} \ge  \mathrm{e}^{2(S+1)\tilde{\Delta}} \mathrm{e}^{-\sum_{k=1}^L\frac{(S+2)x_k^2}{2S}}\right\} \notag\\
&\quad \le \sup_{x^L\,:\, \|x^L\|^2< 2S\Delta}  \E_{p_Z}\left[\mathrm{e}^{-\sum_{k=1}^Lx_kZ_k}\right]\mathrm{e}^{-2(S+1)\tilde{\Delta}}\mathrm{e}^{\sum_{k=1}^L\frac{(S+2)x_k^2}{2S}}\label{MainResultContProofEq7AWGNstepBblock}\\
&\quad = \sup_{x^L\,:\, \|x^L\|^2< 2S\Delta} \mathrm{e}^{-2(S+1)\tilde{\Delta}}\mathrm{e}^{\sum_{k=1}^L\frac{(S+2)x_k^2}{2S}+\frac{x_k^2}{2}}  \notag\\
&\quad <  \mathrm{e}^{-2(S+1)\tilde{\Delta}}\mathrm{e}^{2(S+1)\Delta} \notag\\*
&\quad < \frac{1}{n_m^2}\label{MainResultContProofEq7AWGNblock}
\end{align}
where
\begin{itemize}
\item \eqref{MainResultContProofEq7AWGNstepAblock} follows from the following fact due to the definition of $\imath(\,\cdot\,; \,\cdot\,)$ in~\eqref{infoSpectrumAWGN} and the fact that $\tilde Y_k(1)=Y_k(1)-X_k(1)\sim \mathcal{N}(y_k(1)-x_k(1);0, 1)$ for each $k\in\mathcal{A}$:
\[
    \imath(X_k(1); \tilde Y_k(1))-\imath(X_k(1);  Y_k(1)) = \frac{2S X_k(1) (Y_k(1)-X_k(1)) - (S+2)(X_k(1))^2}{2(S+1)}.
\]
\item \eqref{MainResultContProofEq7AWGNstepBblock} is due to Markov's inequality.
\item \eqref{MainResultContProofEq7AWGNblock} is due to the definitions of~$\Delta$ and $\tilde{\Delta}$ in~\eqref{defDelta} and~\eqref{defDelta+} respectively.
\end{itemize}
Combining~\eqref{MainResultContProofEq6AWGNblock} and~\eqref{MainResultContProofEq7AWGNblock} and using the union bound, we have
\begin{align}
&\Pr\left\{\left.\bigg\{\sum_{k=mL+1}^n \imath(X_k(1); \tilde Y_k(1))  < \log \xi\bigg\}\cap \mathcal{E}_1 \cap \mathcal{E}_2 \right|W=1\right\} \notag\\*
&\quad \le \Pr\left\{\left.\bigg\{\sum_{k=mL+1}^n \imath(X_k(1);  Y_k(1)) - 2S\tilde{\Delta} \left \lceil\frac{|\mathcal{Q}^{(n)}(1)|}{L}\right\rceil < \log \xi\bigg\}\cap \mathcal{E}_1 \cap \mathcal{E}_2 \right|W=1\right\} +\frac{1}{n_m} \label{MainResultContProofEq8AWGNblockStepA} \\*
&\quad \le \Pr\left\{\sum_{k=mL+1}^n \imath(X_k(1);  Y_k(1))  < \log \xi + 2S\tilde{\Delta} (\gamma(\varepsilon_2)+1)\right\} +\frac{1}{n_m} \label{MainResultContProofEq8AWGNblock}
\end{align}
where~\eqref{MainResultContProofEq8AWGNblockStepA} is due to the union bound, the fact that $\mathcal{Q}^{(n)}(1)$ has at most  $\left\lceil\frac{|\mathcal{Q}^{(n)}(1)|}{L}\right\rceil$ blocks of consecutive mismatched positions (only the last block may have length other than~$L$), and the fact that~\eqref{MainResultContProofEq7AWGNblock} holds if~$L$ is replaced with any natural number $L^*\le L$; and~\eqref{MainResultContProofEq8AWGNblock} follows from the definition of $\mathcal{E}_1$ in~\eqref{defE1block}. The first term in~\eqref{MainResultContProofEq8AWGNblock} can be bounded by standard procedures which will be elaborated later. In order to bound the second term in~\eqref{MainResultContProofEq5AWGNblock}, we use Lemma~\ref{lemmaFeinsteinBlock} (Lemma~\ref{lemmaFeinstein} suffices for the case~$L=1$) to obtain
\begin{align}
&\Pr\left\{\left.\bigg\{\sum_{k=mL+1}^n \imath(X_k(2); \tilde Y_k(1)) \ge \log \xi\bigg\} \cap \mathcal{E}_1 \cap \mathcal{E}_2\right|W=1\right\}
 \le \frac{2}{M}\,\mathrm{e}^{-(\log \xi - \log M)}\times \big(n_m(S+1)^{L/2}\big)^{\gamma(\varepsilon_2)+1}.
\label{MainResultContProofEq9AWGNblock}
\end{align}
Consequently, it follows from \eqref{MainResultContProofEq2AWGNblock}, \eqref{MainResultContProofEq5AWGNblock}, \eqref{MainResultContProofEq8AWGNblock} and~\eqref{MainResultContProofEq9AWGNblock} that
\begin{align}
 \Pr\left\{\varphi\big(\tilde Y^n(W)\big)\ne W\right\} &\le  \Pr\left\{\sum_{k=mL+1}^n \imath(X_k(1);  Y_k(1))  < \log \xi + 2S\tilde{\Delta} \lceil \gamma(\varepsilon_2)+1 \rceil\right\} \notag\\*
  &\quad +2\mathrm{e}^{-\big(\log \xi-\log M - (\gamma(\varepsilon_2)+1)\log(n_m(S+1)^{L/2}) \big)} + \varepsilon_2+ \frac{1}{\sqrt{n_m}}+\frac{1}{n_m}.\label{MainResultContProofEq12AWGNblock}
\end{align}
The remainder of the proof follows from standard steps, outlined below for the sake of completeness. Let $\mu=\frac{1}{2}\log (1+S)$, $\sigma^2=\frac{S}{S+1}>0$ and $T<\infty$ denote the mean, the variance and the third absolute moment of $\imath(X; Y)$ respectively, where the finiteness of~$T$ is due to~\eqref{infoSpectrumAWGN} and the fact that $|S|\le P$. Choose
 \begin{align}
 \log \xi \triangleq n_m\mu + \sqrt{n_m \sigma^2}\,\Phi^{-1}\left(\varepsilon_1 -\frac{T}{\sigma^3\sqrt{n_m}}-\frac{4}{\sqrt{n_m}}\right) -2S\tilde{\Delta}(\gamma(\varepsilon_2)+1). \label{defXiAWGNblock}
 \end{align}
It then follows from Berry-Ess\'een theorem~\cite{KorolevShevtsova10}, i.e.,
\begin{align*}
\left|\Pr\left\{\frac{\sum_{k=1}^n \imath(X_k; Y_k) - n\mu}{\sqrt{n\sigma^2}}\le a\right\}-\Phi(a)\right|\le \frac{T}{\sigma^3\sqrt{n}} 
\end{align*}
 for all $a\in\mathbb{R}$, that
\begin{align}
\Pr\left\{\sum_{k=mL+1}^n \imath(X_k(1); Y_k(1))  < \log \xi + 2S\tilde{\Delta}(\gamma(\varepsilon_2)+1) \right\}\le \varepsilon_1 - \frac{4}{\sqrt{n_m}}. \label{eq9ErrorProbAWGNblock}
\end{align}
 In order to bound the second term in~\eqref{MainResultContProofEq12AWGNblock}, we choose
 \begin{align}
 \log M &\triangleq \left\lfloor  \log \xi -  (\gamma(\varepsilon_2)+1)\left(\frac{L}{2}\log(S+1) + \log n_m\right) -\log\sqrt{n_m}\right\rfloor  \label{defMAWGN*block}\\*
 & \ge  \log \xi -   (\gamma(\varepsilon_2)+1)\left(\frac{L}{2}\log(S+1) + \log n_m\right)- \log\sqrt{n_m} -1.\label{defMAWGNblock}
 \end{align}
 Consequently, \eqref{st2ThmMainResultAWGNblock} follows from~\eqref{MainResultContProofEq12AWGNblock}, \eqref{eq9ErrorProbAWGNblock} and~\eqref{defMAWGN*block}, and~\eqref{st1ThmMainResultAWGNblock} follows from~\eqref{defXiAWGNblock} and~\eqref{defMAWGNblock}. In addition, \eqref{st3ThmMainResultAWGNblock} follows from~\eqref{MainResultContProofEq2AWGNstepAblock}.

 \section{Numerical Results} \label{sectionNumerical}
 In this section, we numerically compare the performance of our analyzed save-and-transmit with the state-of-the-art save-and-transmit in~\cite{FTO17} under the following two cases: The i.i.d.\ energy arrival case with $L=1$, and the block energy arrival case with $L=\lceil\sqrt{n}\,\rceil$. In both cases, we assume that $\E[E^2]= 3(\E[E])^2$. An example for~$E$ is $E= U^2$ where $U\sim \mathcal{N}(u; 0, P)$. The major difference between the two save-and-transmit strategies is that the former one uses a transmit power~$S$ strictly less than the battery recharge rate~$P$ while the latter one always assumes $S=P$. The difference in transmitting power results in different achievable rates as shown in the rest of the section.
 \subsection{Case $L=1$}
 \begin{figure}[!t]
\centering
   \subfigure[SNR = 25 dB]{
        \includegraphics[width=3 in]{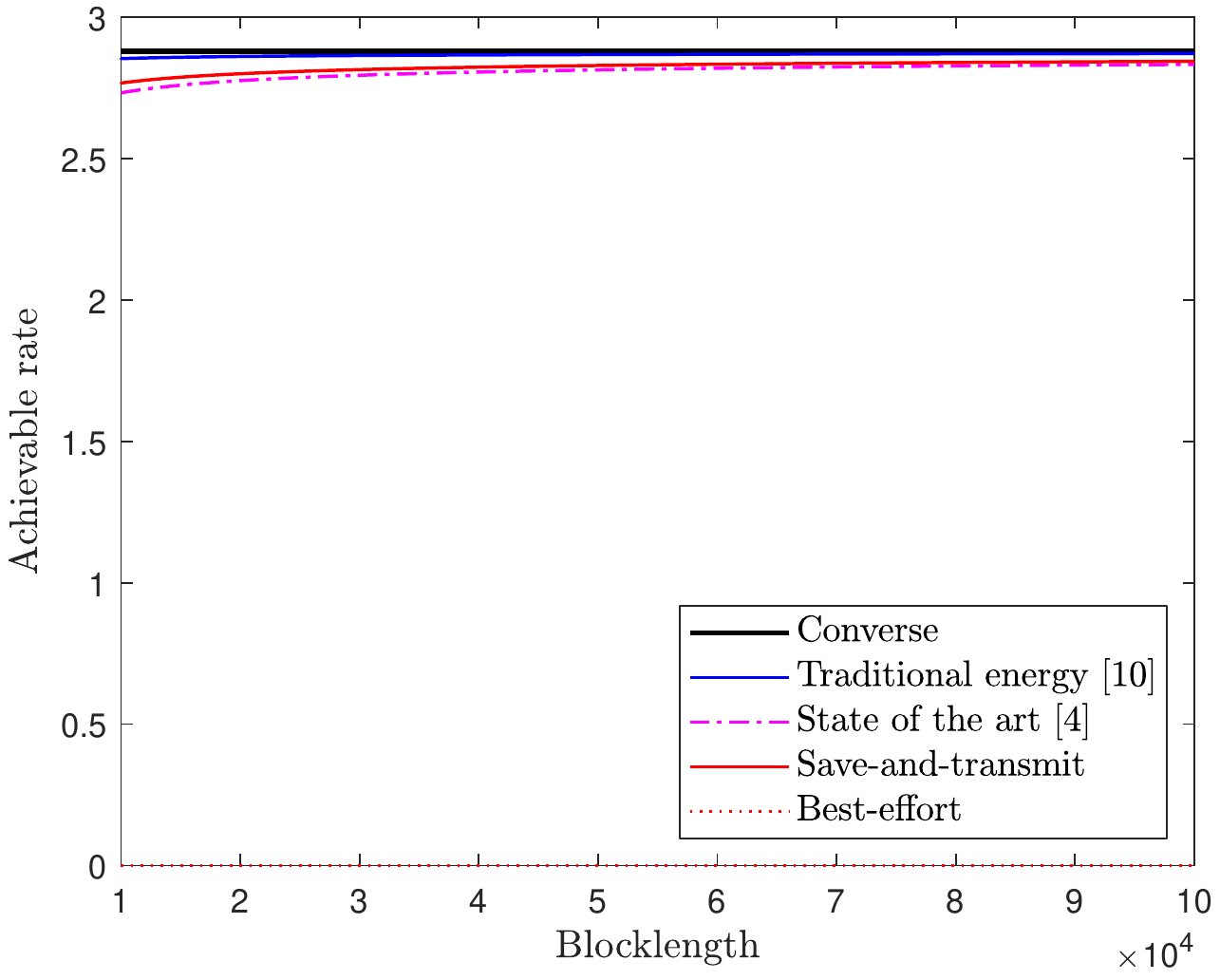}

   }
\subfigure[SNR = 0 dB]{
        \includegraphics[width=3 in]{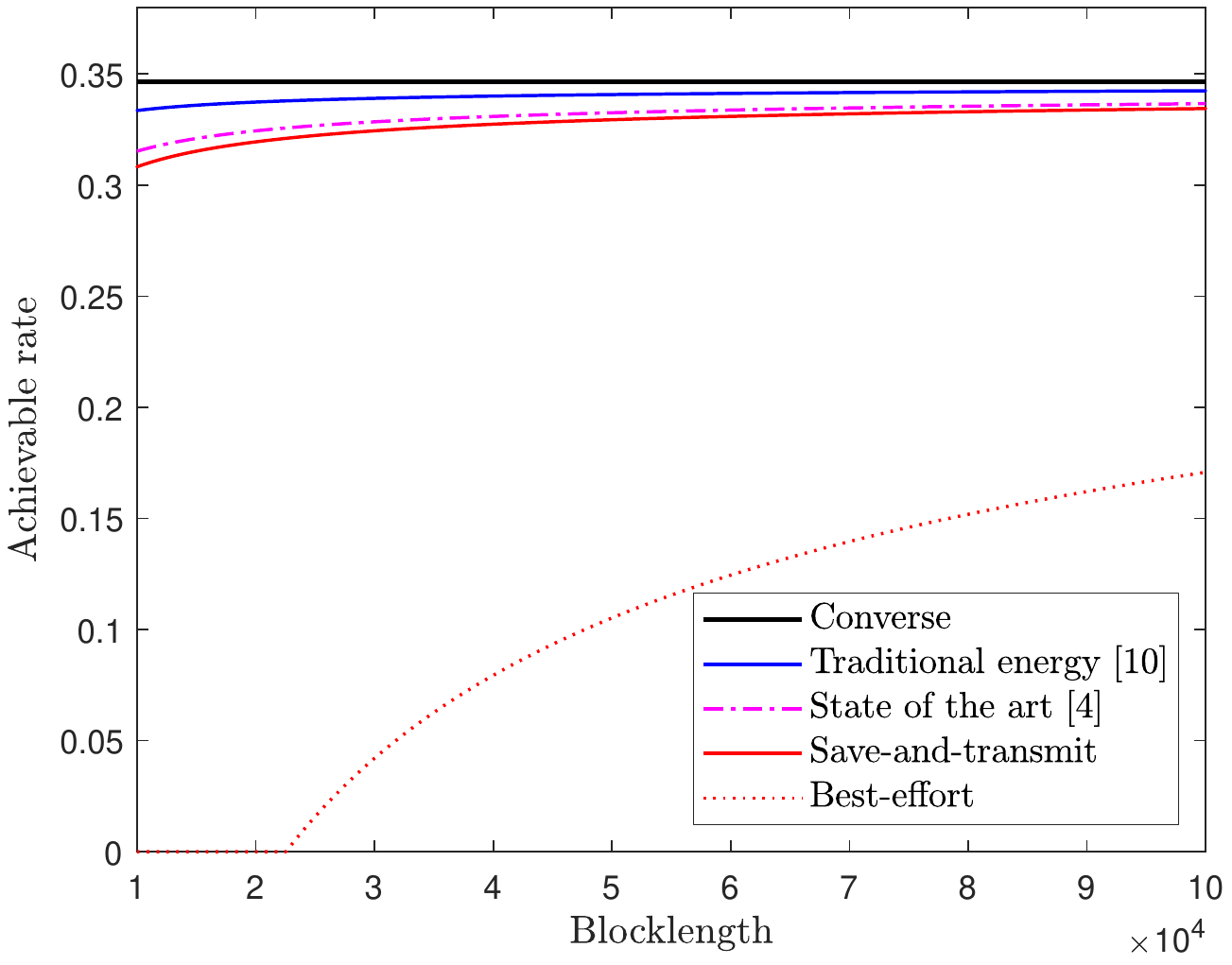}

   }
\caption{Achievable rates for save-and-transmit, best-effort and the state of the art~\cite{FTO17} for $L=1$ where $\varepsilon_1=\varepsilon_2=0.01$.}
\label{figure1}
\end{figure}
Figure~\ref{figure1}(a) plots the achievable rate up to the $\Theta(1/\sqrt{n})$ term of our analyzed save-and-transmit scheme, our analyzed best-effort scheme and the state-of-the-art save-and-transmit~\cite[Th.~1]{FTO17} according to~\eqref{st1CorollaryMainResultAWGNsaving}, \eqref{st1CorollaryMainResultAWGN} and~\eqref{existingBest} respectively for the high SNR (i.e., battery recharge rate) regime where $P=25$ dB, $\E[X^2]=3P^2$, and $\varepsilon_1=\varepsilon_2=0.01$. Note that best-effort does not achieve a positive rate in this regime because the magnitude of the backoff term $-\sqrt{\frac{(\lambda_1 + \lambda_2\log n)(\E[E^2]+3P^2)\log\frac{1}{\varepsilon_2}}{P(P+1)}}\times\frac{1}{\sqrt{n}}$ is larger than the capacity $\frac{1}{2}\log(1+P)$ for large~$P$. In addition, we compare in Figure~\ref{figure1}(b) the three schemes for the low SNR regime $P=0$ dB. For the high SNR regime, Figure~\ref{figure1}(a) shows that save-and-transmit outperforms the state of the art at reasonable values of the blocklength. On the other hand, the state of the art outperforms save-and-transmit for the low SNR regime as shown in Figure~\ref{figure1}(b). The two plots in Figure~\ref{figure1} agree with Remark~\ref{remarkCorollarySaving} and Remark~\ref{remarkBestEffort}. To demonstrate the effect of EH constraints~\eqref{eqn:eh} on the AWGN channel, we also plot the following maximum achievable rate up to the $\Theta(1/\sqrt{n})$ term~\cite[Th.~54, Eq.~(294)]{Pol10} when the EH constraints are replaced with the conventional power constraint $\Pr\{\frac{1}{n}\sum_{k=1}^n X_k^2 \le nP\}=1$:
\begin{align}
\frac{1}{2}\log(1+P)+\sqrt{\frac{P(P+2)}{2(P+1)^2}}\Phi^{-1}(\varepsilon_1+\varepsilon_2). \label{rateNoEH}
\end{align}

  \subsection{Case $L=\lceil\sqrt{n}\,\rceil$}
  \begin{figure}[!t]
\centering
   \subfigure[SNR = 25 dB]{
        \includegraphics[width=3 in]{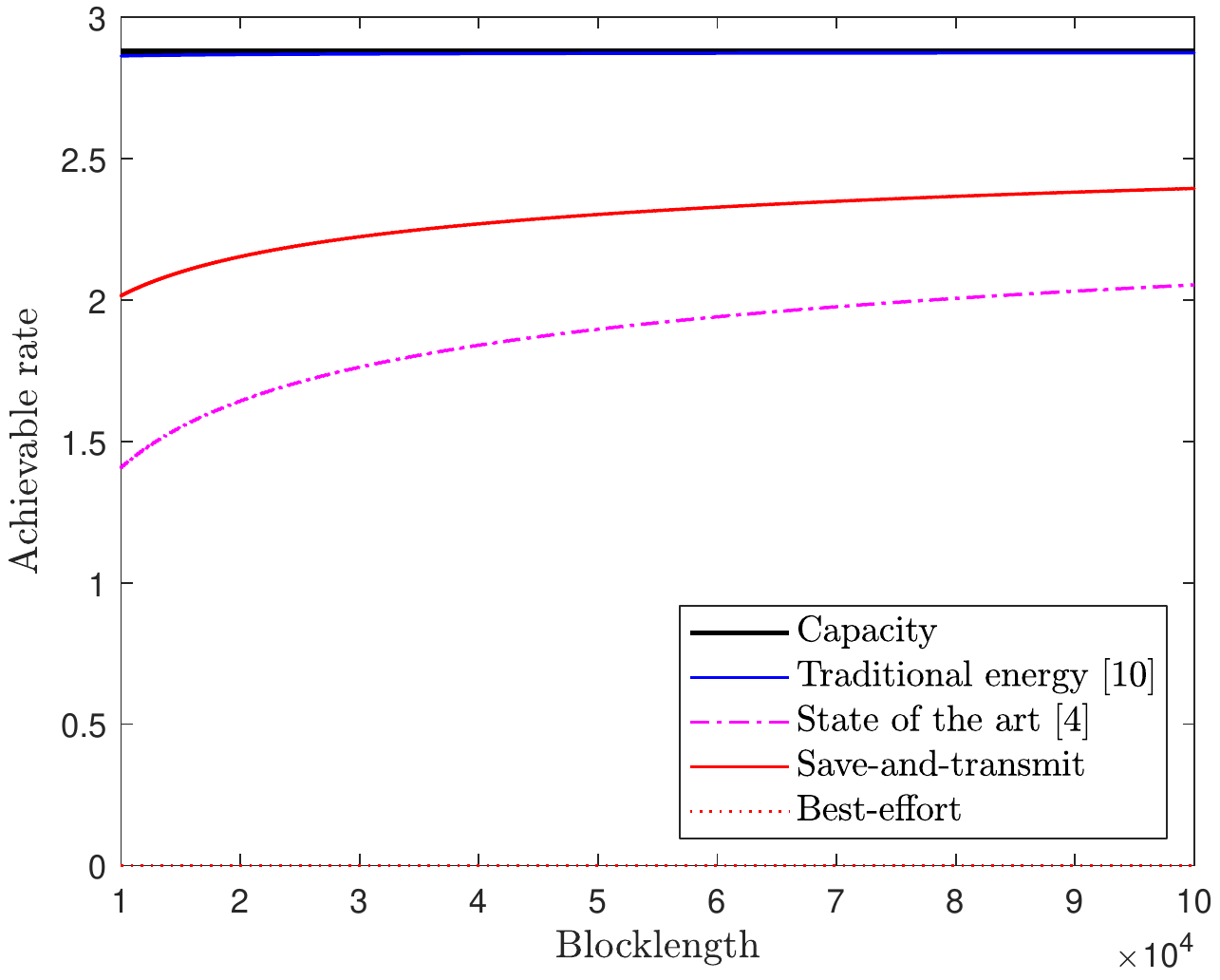}

   }
\subfigure[SNR = 0 dB]{
        \includegraphics[width=3 in]{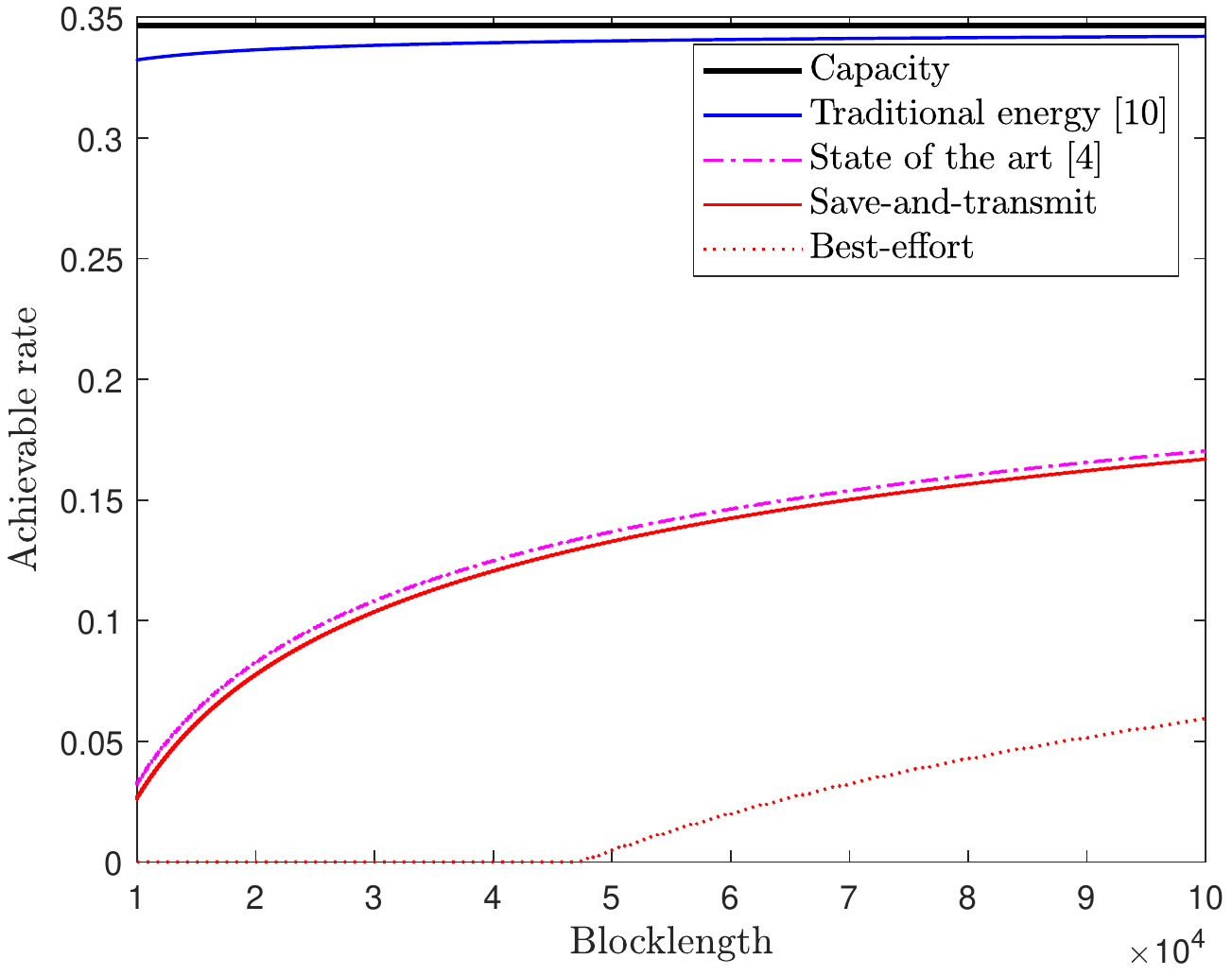}

   }
\caption{Achievable rates for save-and-transmit, best-effort and the state of the art~\cite{FTO17} for $L=\lceil\sqrt{n}\,\rceil$ and $\varepsilon=0.01$.}
\label{figure2}
\end{figure}
Figure~\ref{figure2}(a) plots the achievable rate up to the $\Theta(\sqrt{L/n})$ term of our analyzed save-and-transmit scheme, our analyzed best-effort scheme and the state-of-the-art save-and-transmit~\cite[Th.~1]{FTO17} according to~\eqref{st1ThmBlockSave}, \eqref{st1ThmBlockBestEffort} and~\eqref{existingBestBlock} respectively for the high SNR regime $P=25$ dB and $\E[X^2]=3P^2$ and for $\varepsilon=0.01$. Note that best-effort does not achieve a positive rate in this regime because the magnitude of the backoff term $ -\sqrt{\frac{\big(2P\log 2+\frac{1}{2}\log(1+P)\big)\E[E^2]\log\frac{1}{\varepsilon}}{P(P+1)}}\times \sqrt{\frac{L}{n}}$ is larger than the capacity $\frac{1}{2}\log(1+P)$ for large~$P$. In addition, we compare in Figure~\ref{figure2}(b) the three schemes for the low SNR regime $P=0$ dB. For the high SNR regime, Figure~\ref{figure2}(a) shows that save-and-transmit outperforms the state of the art at reasonable values of the blocklength. On the other hand, the state of the art outperforms save-and-transmit for the low SNR regime as shown in Figure~\ref{figure2}(b). The two plots in Figure~\ref{figure2} agree with Remark~\ref{remarkCorollarySavingBlock} and Remark~\ref{remarkBestEffortBlock}. To demonstrate the effect of EH constraints~\eqref{eqn:eh} on the AWGN channel, we also plot the maximum achievable rate~\eqref{rateNoEH} up to the $\Theta(1/\sqrt{n})$ term with $\varepsilon_1+\varepsilon_2=0.01$ when the EH constraints are replaced by $\Pr\{\sum_{k=1}^n X_k^2 \le nP\}=1$.

\subsection{Impact of SNR}
 \begin{figure}[!t]
\centering
   \subfigure[$L = 1$ and $\varepsilon_1=\varepsilon_2=0.01$]{
        \includegraphics[width=3 in]{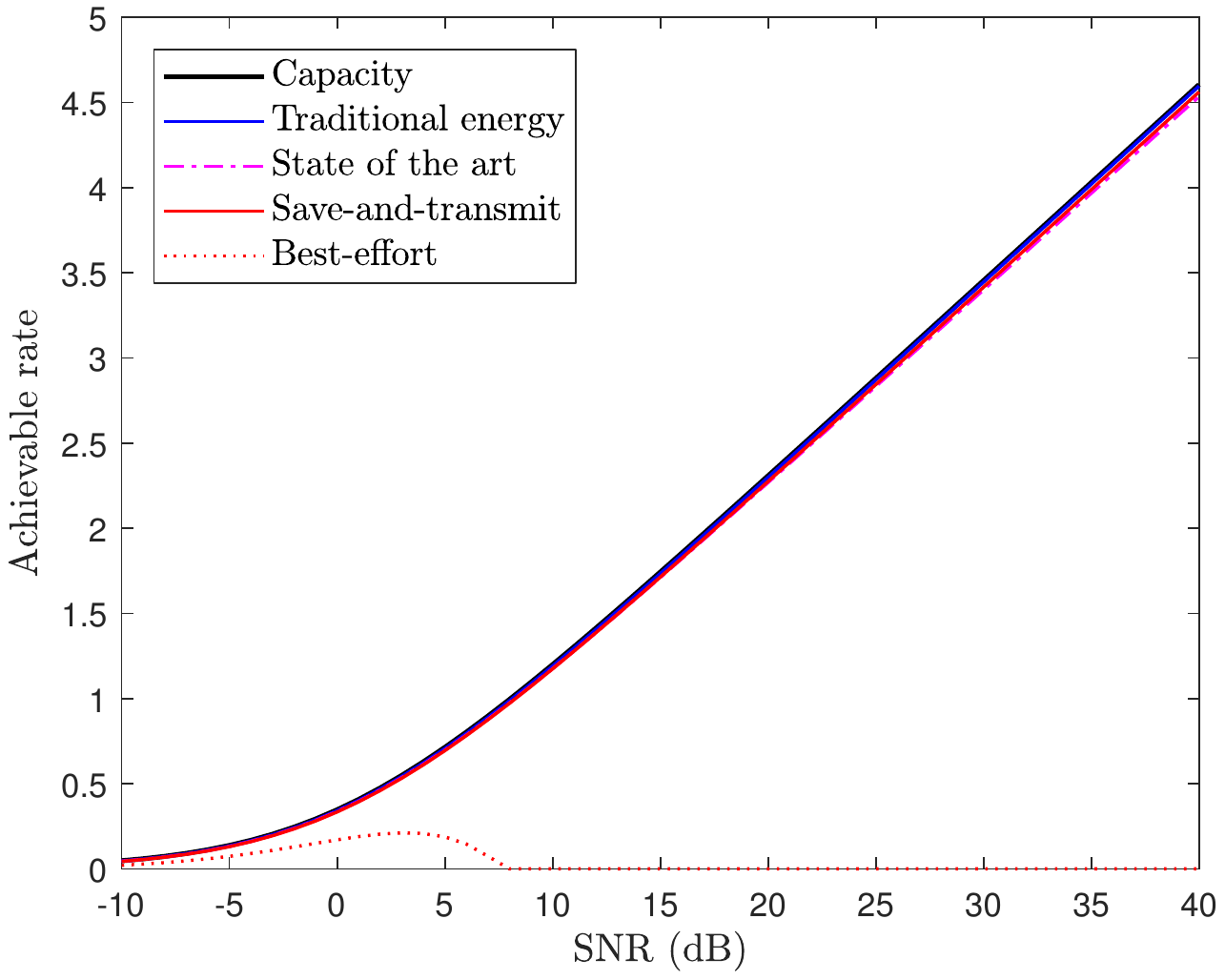}

   }
\subfigure[$L=\lceil\sqrt{n}\,\rceil$ and $\varepsilon=0.01$]{
        \includegraphics[width=3 in]{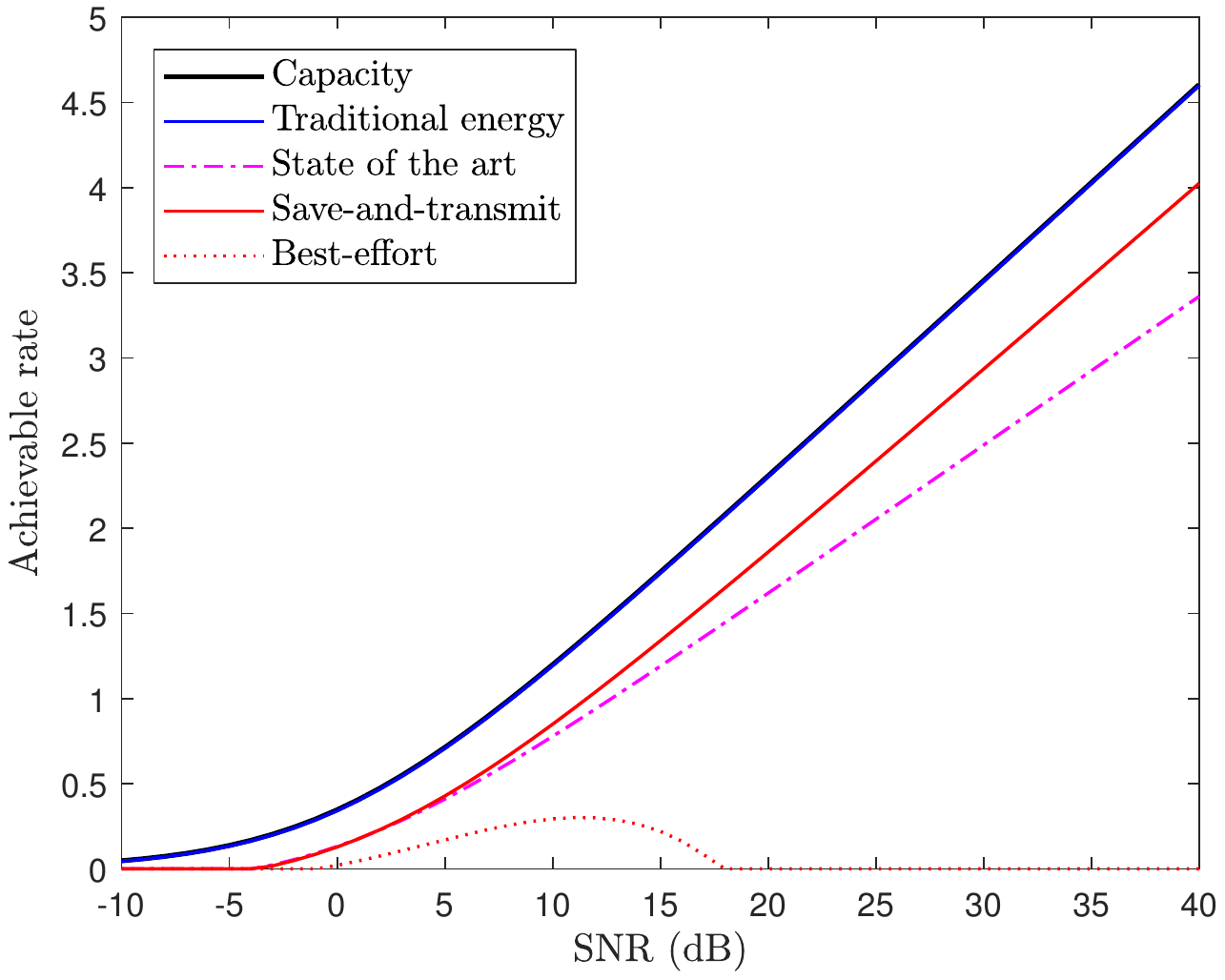}

   }
\caption{Achievable rates for save-and-transmit, best-effort and the state of the art~\cite{FTO17} for $n=10^5$.}
\label{figure3}
\end{figure}
In order to illustrate how the SNR impacts the performance of the save-and-transmit, best-effort and the state of the art, we plot in Figure~\ref{figure3}(a) their achievable rates at a fixed blocklength up to the $\Theta(1/\sqrt{n})$ term against SNR for $L=1$, $n=10^5$ and $\varepsilon_1=\varepsilon_2=0.01$. Similarly, we plot in Figure~\ref{figure3}(b) their achievable rates up to the $\Theta(\sqrt{L/n})$ term against SNR for $L=\lceil\sqrt{n}\,\rceil$, $n=10^5$ and $\varepsilon=0.01$. For $L=1$, save-and-transmit and the state of the art have similar performance. In contrast, for $L=\lceil\sqrt{n}\,\rceil$, save-and-transmit outperforms the state of the art when the SNR is larger than 5 dB. For both cases $L=1$ and $L=\lceil\sqrt{n}\,\rceil$, best-effort achieves a positive rate only within a range of SNRs. Therefore, recalling the major difference between save-and-transmit and the state of the art explained at the beginning of this section, we conclude that allowing the transmit power to be strictly less than the SNR (i.e., battery recharge rate) can be beneficial for the block energy arrival case.

\section{Concluding Remarks} \label{conclusion}
In this paper, we have studied in the finite blocklength regime the save-and-transmit scheme with a saving phase of arbitrary length~$m$ over the AWGN EH channel, and also the best-effort scheme through setting $m=0$. A new non-asymptotic achievable rate is obtained for save-and-transmit, which directly implies a new non-asymptotic achievable rate for best-effort. The non-asymptotic result implies that the save-and-transmit scheme achieves the optimal second-order scaling $-O\big(\frac{1}{\sqrt{n}}\big)$, and that the best-effort scheme achieves the second-order scaling $-O\big(\sqrt{\frac{\log n}{n}}\big)$. The achievable rates for the schemes are extended to the block energy arrival model where $L=o(n)$, and are shown to achieve the second-order scalings $-O\Big(\sqrt{\frac{\max\{\log n, L\}}{n}}\Big)$ and $-O\Big(\sqrt{\frac{L}{n}}\Big)$ respectively. In particular, both analyzed schemes achieve the optimal scaling $-O\Big(\sqrt{\frac{L}{n}}\Big)$ for the case $L=\omega(\log n)$. Compared to the state-of-the-art save-and-transmit scheme~\cite{FTO17}, our save-and-transmit has a better finite-blocklength performance for sufficiently large~$P$.

For the simplest case $L=1$, the best-effort scheme does not achieve the optimal scaling $-O\big(\frac{1}{\sqrt{n}}\big)$. A straightforward verification by MATLAB reveals that under the assumption $E= U^2$ where $U\sim\mathcal{N}(u; 0, 1)$, the average number of mismatched positions for a best-effort scheme is of the order $o(\sqrt{n})$. A future direction may involve improving the second-order scaling $-O\Big(\sqrt{\frac{\log n}{n}}\Big)$ for $L=1$ for best-effort schemes by possibly proving a sharper probability bound than~\eqref{stLemmaMismatch} in Lemma~\ref{lemmaSavingMismatch}. Another interesting direction is to tighten the existing non-asymptotic upper bound for a general coding scheme presented in~\cite[Th.~1]{FTO17}, which states that the second-order term is bounded above by $\frac{\sqrt{2P^2+\E[E^2]}}{2(P+1)}\Phi^{-1}(\varepsilon)\times \sqrt{\frac{L}{n}}$. The upper bound is potentially loose because it considers only the last EH constraint $\sum_{i=1}^n X_k^2 \le \sum_{i=1}^n E_k$ rather than the~$n$ EH constraints in~\eqref{eqn:eh}. Last but not least, a natural extension of this work is to explore non-asymptotic achievable rates for EH channels with finite battery~\cite{SNOzgur2016,MaoHassibi2017}.

\appendices

\section{Proofs of Lemma~\ref{lemmaSavingMismatch} and Lemma~\ref{lemmaMismatchBlock}}\label{appendixAblock}
Since save-and-transmit defined in Section~\ref{subsecSaveTransmit} is a special case of blockwise save-and-transmit defined in Section~\ref{subsecSaveTransmitBlock} with $L=1$ and Lemma~\ref{lemmaSavingMismatch} is a special case of Lemma~\ref{lemmaMismatchBlock} with $L=1$, it suffices to prove Lemma~\ref{lemmaMismatchBlock}.

 Fix an~$n\in\mathbb{N}$, a natural number $L<n$ and a $\rho\in(0,1)$ that satisfies~\eqref{sufficientlyLargeN},
 and fix a blockwise save-and-transmit $(n, M)$-EH code as described in Section~\ref{subsecSaveTransmitBlock}.
 Let $p_X$ be as defined in~\eqref{defDistPX} where~$S$ is as defined in~\eqref{defS}. Define $p_{\hat E}$ to be the distribution of $\hat E\triangleq \sum_{j=1}^L E_j= L E$ where~$E$ satisfies~\eqref{defEnergyP}, and define $p_{\hat X}$ to be the distribution of $\hat X\triangleq \sqrt{\sum_{j=1}^{L} X_j^2}$.
In this proof, all the probability, expectation and variance terms are evaluated according to $p_{\hat X^\infty}p_{\hat E^\infty}$ where $p_{\hat X^\infty}=\prod_{k=1}^\infty p_{\hat X_k}$ and $p_{\hat E^\infty}=\prod_{k=1}^\infty p_{\hat E_k}$ denote the infinite product distributions of $p_{\hat X}$ and $p_{\hat E}$ respectively. Consider the Markov process $\left\{\sum_{i=1}^m \hat E_{i}+\sum_{i=m+1}^{m+k} \big(\hat E_{i}- \hat X_i^2\big)\right\}_{k=1}^{\tau(m)}$ where~$m$ is an arbitrary non-negative integer and~$\tau(m)$ is the stopping time when the value of the Markov process hits any $a< 0$. By definition of~$\tau(m)$, we have
\begin{align*}
\Pr\left\{\left. \sum_{i=1}^m \hat E_{i}+\sum_{i=m+1}^{m+\tau(m)} \big(\hat E_{i}- \hat X_i^2\big) <0 \,\right| \tau(m)<\infty \right\}=1
\end{align*}
and
\begin{align}
\Pr\left\{\tau(m)=\infty \right\}= \Pr\left\{\bigcap_{k=1}^\infty \left\{\sum_{i=1}^m \hat E_{i}+\sum_{i=m+1}^{m+k} \big(\hat E_{i}- \hat X_i^2\big)\ge 0\right\}\right\} \label{appendixLemma1Eq0-Block}
\end{align}
for each $m\in\mathbb{Z}_+$ where $\Pr\left\{\tau(m)=\infty \right\}$ denotes the escape probability.

In order to show~\eqref{stLemmaMismatchBlock}, we first fix a~$\gamma\in\mathbb{N}$ and let $\tau_1, \tau_2, \ldots, \tau_\gamma$ be~$\gamma$ independent copies of~$\tau(1)$. Due to the construction of the blockwise save-and-transmit scheme with a saving phase of length~$mL$, energy is saved but not consumed during the saving phase and each block of~$L$ consecutive mismatch events. Therefore, $\tau(m)$ serves as a lower bound on the number length-$L$ blocks between the first length-$L$ block in the transmission phase and the first block of mismatch events (excluding one block) and~$\tau(1)$ serves as a lower bound on the number of blocks between two blocks of mismatch events (excluding one block). Fix any~$w\in\mathcal{W}$ and consider
\begin{align}
\Pr\left\{\left.|\mathcal{Q}^{(n)}(w)| \ge L\gamma+1\right|W=w\right\}
&=
\Pr\left\{\left.\text{$\mathcal{Q}^{(n)}(w)$ contains at least $\gamma + 1$ blocks of mismatch events}\right|W=w\right\} \notag\\
&\le \Pr\left\{\{\tau(m)<\infty\}\cap \bigcap_{k=1}^\gamma \{\tau_k<\infty\}\right\} \notag\\
& = \Pr\{\tau(m)<\infty\}\left(\Pr\{\tau(1)<\infty\}\right)^\gamma. \label{appendixLemma1Eq0Block}
\end{align}
In order to obtain an upper bound on
$
\Pr\left\{\tau(m)<\infty \right\}$,
we first construct the following sequence denoted by $\{\hat B_k\}_{k=1}^{\infty}$. For each $k\in\mathbb{N}$, define $\hat B_k$ recursively as
\begin{align}
\hat B_k\triangleq
&\begin{cases}
\hat E_1 & \text{if $k=1$ and $m\ge 1$,}\\
\hat E_1-\hat X_1^2 & \text{if $k=1$ and $m=0$,} \\
\hat B_{k-1} +  \hat E_k & \text{if $k\in\{2, 3, \ldots, m\}$,}\\
\hat B_{k-1} +\hat  E_k - \hat X_{k}^2 & \text{if $k\ge m+1$ and $\hat B_{k-1} \ge 0$,}\\
\hat B_{k-1} & \text{if $k\ge m+1$ and $\hat B_{k-1} < 0$.}
\end{cases} \label{defBkBlock}
\end{align}
By inspecting~\eqref{defBkBlock}, we have
\begin{align}
\{\hat B_{\infty}< 0\}&=\bigcup_{k=1}^\infty\left\{\sum_{i=1}^m \hat E_i+\sum_{i=m+1}^{m+k} (\hat E_i - \hat X_i^2)< 0\right\} \label{appendixLemma1Eq1Block}\\
&=\{\tau(m)<\infty\},  \label{appendixLemma1Eq1*Block}
\end{align}
where each term in the union in~\eqref{appendixLemma1Eq1Block} characterizes the event that the accumulated energy collected during the first $m+k$ energy blocks is insufficient to
output the desired codeword symbols from block~$m+1$ to block~$m+k$ during the transmission phase.
It remains to obtain an upper bound on $\Pr\{\hat B_{\infty}< 0\}$.
To this end, we first define for each $k\in\mathbb{N}$
 \begin{align}
 \hat U_k &\triangleq
 \begin{cases}
  \hat B_1 & \text{if $k=1$,}\\
 \hat B_k-\hat B_{k-1} & \text{otherwise}
 \end{cases}   \notag\\ 
 &=
 \begin{cases}
\hat B_1 & \text{if $k=1$,}\\
\hat E_k -\hat  X_k^2 & \text{if $k\in\{2, 3, \ldots\}$ and $\hat B_{k-1} \ge 0$,}\\
0 & \text{if $k\in\{2, 3, \ldots\}$ and $\hat B_{k-1} < 0$}
\end{cases}  \label{defUk*Block}
 \end{align}
 where  \eqref{defUk*Block} follows from~\eqref{defBkBlock}. It then follows from~\eqref{appendixLemma1Eq1Block} and~\eqref{defUk*Block} that $ \{\hat B_{\infty}< 0\} =  \left\{\sum_{k=1}^\infty \hat U_k< 0\right\}$, hence
 \begin{align}
 \Pr\{\hat B_{\infty}< 0\} =  \Pr\left\{\sum_{k=1}^\infty \hat U_k< 0\right\}. \label{appendixLemma1Eq1**Block}
 \end{align}
 Following~\eqref{appendixLemma1Eq1**Block}, we consider the chain of inequalities below for any $t>0$:
 \begin{align}
 \Pr\left\{\sum_{k=1}^{\infty}\hat U_k< 0\right\}
 &= \Pr\left\{\mathrm{e}^{-t\sum_{k=1}^{\infty}\hat U_k} > 1\right\}\notag\\*
 &\le \E\left[\mathrm{e}^{-t\sum_{k=1}^{\infty}\hat U_k}\right] \label{appendixLemma1Eq2Block}
 \end{align}
 where the inequality follows from Markov's inequality. In order to simplify the RHS of \eqref{appendixLemma1Eq2Block}, we use the convention $\hat E^0=\hat X^0=\hat U^0=0$ (useful only when~$m=0$) and consider the following chain of inequalities for each $i\in\{m+1, m+2, \ldots\}$:
 \begin{align}
 \E\left[\mathrm{e}^{-t\sum_{k=1}^{i}\hat U_k}\right]
 & =  \E\left[ \E\left[\left.\mathrm{e}^{-t\sum_{k=1}^{i}\hat U_k}\right| \hat U^{i-1} \right] \right] \notag\\
 & =   \E\left[ \E\left[ \E\left[\left.\mathrm{e}^{-t\hat U_{i}}\right| \hat  U^{i-1}\right]\left.\mathrm{e}^{-t\sum_{k=1}^{i-1}\hat U_k}\right| \hat U^{i-1} \right] \right]  \notag\\
 &\le \E\left[ \E\left[\max\left\{\E\left[\left. \mathrm{e}^{-t(\hat E_{i} -\hat X_i^2)}\right|\hat U^{i-1}\right] , 1\right\} \times \mathrm{e}^{-t\sum_{k=1}^{i-1}\hat U_k} \Big| \hat U^{i-1}\right]                 \right] \label{appendixLemma1Eq3StepABlock} \\*
 &=\max\left\{\E\left[ \mathrm{e}^{-t(\hat E_{i} - \hat X_i^2)}\right] , 1\right\}  \E\left[\mathrm{e}^{-t\sum_{k=1}^{i-1}\hat U_k}\right]\label{appendixLemma1Eq3Block}
 \end{align}
 where \eqref{appendixLemma1Eq3StepABlock} is due to~\eqref{defUk*Block};
\eqref{appendixLemma1Eq3Block} follows from the independence between $(\hat E_{i}, \hat X_{i})$ and $\hat U^{i-1}$ due to the independence between $(\hat E_{i}, \hat X_{i})$ and $(\hat E^{i-1}, \hat X^{i-1})$.

  Combining~\eqref{appendixLemma1Eq1**Block}, \eqref{appendixLemma1Eq2Block} and \eqref{appendixLemma1Eq3Block}, we have
 \begin{align}
  \Pr\left\{\hat B_{\infty}< 0\right\} &\le \E\left[\mathrm{e}^{-t \sum_{k=1}^m \hat U_k}\right]  \max\left\{ \left(\E\left[ \mathrm{e}^{-t(\hat E - \hat X^2)}\right]\right)^{\infty},1\right\} \notag\\
  &=  \left(\E\left[\mathrm{e}^{-t \hat E}\right]\right)^m  \max\left\{ \left(\E\left[ \mathrm{e}^{-t(\hat E - \hat X^2)}\right]\right)^{\infty},1\right\},  \label{appendixLemma1Eq4Block}
 \end{align}
 which together with the definitions of~$\hat E$ and~$\hat X^2$ implies that
 \begin{align}
  \Pr\left\{\hat B_{\infty}< 0\right\} \le \left(\E\left[\mathrm{e}^{-t L  E}\right] \right)^m \max\left\{ \left(\E\left[ \mathrm{e}^{-t(L  E - \sum_{j=1}^L  X_j^2)}\right]\right)^{\infty},1\right\}. \label{appendixLemma1Eq4Block}
 \end{align}
  In order to simplify the RHS of~\eqref{appendixLemma1Eq4Block}, we use the following two facts, whose proofs can be found in~\cite[Appendix]{FTY15}:
 For any $y\ge 0$,
\begin{equation}
1+y\le \mathrm{e}^y \le 1+ y+ \frac{y^2 \mathrm{e}^y}{2} \label{propositionStatement1Block}
\end{equation}
and
\begin{equation}
1-y \le \mathrm{e}^{-y} \le 1 - y+ \frac{y^2}{2}\,. \label{propositionStatement2Block}
\end{equation}
Let $t>0$ be the positive solution of the quadratic equation
\begin{align}
t= \frac{2(P-S)}{L\E[E^2]+3S^2(1+63St)}. \label{sufficientlySmallTBlock}
\end{align}
Straightforward calculations reveal that
\begin{align}
t&=\frac{-(L\E[E^2]+3S^2)+\sqrt{(L\E[E^2]+3S^2)^2+1512S^3(P-S)}}{378S^3}\notag\\
&\le \frac{\sqrt{42(P-S)}}{63\sqrt{S^3}}\label{sufficientlySmallT***stepABlock}\\
&=\frac{\sqrt{42\rho P}}{63 \sqrt{S^3}} \notag\\
&<\frac{1}{6S} \label{sufficientlySmallT***Block}
\end{align}
where \eqref{sufficientlySmallT***stepABlock} is due to the fact that $\sqrt{a+b}\le \sqrt{a}+\sqrt{b}$ for all $(a,b)\in\mathbb{R}_+^2$; \eqref{sufficientlySmallT***Block} is due to \eqref{sufficientlyLargeN} and~\eqref{defS}.
 Using the definition of $p_X$ in~\eqref{defDistPX} and~\eqref{sufficientlySmallT***Block}, we have
\begin{align}
\E\left[X^4 \mathrm{e}^{tX^2}\right] = \frac{3S^2}{(1-2St)^{5/2}} <\infty. \label{sufficientlySmallT*Block}
\end{align}
In addition, using~\eqref{sufficientlySmallT***Block} and straightforward algebra, we obtain
\begin{align}
\frac{1}{(1-2St)}\le 1+3St \label{straightForwardFact1Block}
\end{align}
and
\begin{align}
(1+3St)^{5/2}&\le (1+3St)^{3}\notag\\
&\le 1+9St+27S^2t^2+27S^3t^3\notag\\
&\le 1+63St. \label{straightForwardFact2Block}
\end{align}
Following~\eqref{appendixLemma1Eq4Block}, we use the two facts~\eqref{propositionStatement1Block} and \eqref{propositionStatement2Block} to obtain
\begin{align}
\E\left[\mathrm{e}^{-tLE}\right] \le 1-tLP + \frac{t^2L^2\E[E^2]}{2} \le \mathrm{e}^{-tLP + \frac{t^2L^2\E[E^2]}{2}}  \label{appendixLemma1Eq7Block}
\end{align}
and
\begin{align*}
\E\left[\mathrm{e}^{tX^2}\right] \le 1+tS + \frac{t^2\E[X^4 \mathrm{e}^{tX^2}]}{2} \le \mathrm{e}^{tS + \frac{t^2\E[X^4 \mathrm{e}^{tX^2}]}{2}},
\end{align*}
which implies that
\begin{align}
\E\left[\mathrm{e}^{-tLE}\right]\E\left[\mathrm{e}^{t\sum_{j=1}^L  X_j^2}\right]&\le \mathrm{e}^{-tL(P-S) + \frac{t^2L}{2}\left(L\E[E^2]+\E[X^4 \mathrm{e}^{tX^2}]\right)} \notag\\
&\le  1   \label{appendixLemma1Eq8Block}
\end{align}
where~\eqref{appendixLemma1Eq8Block} follows from the fact due to~\eqref{sufficientlySmallTBlock}, \eqref{sufficientlySmallT*Block}, \eqref{straightForwardFact1Block} and~\eqref{straightForwardFact2Block} that
\begin{align*}
t \le \frac{2(P-S)}{L\E[E^2]+\E\left[X^4 \mathrm{e}^{tX^2}\right]}.
\end{align*}
Using~\eqref{appendixLemma1Eq4Block},  \eqref{appendixLemma1Eq7Block} and~\eqref{appendixLemma1Eq8Block}, we obtain
\begin{align}
 \Pr\left\{\hat B_{\infty}< 0\right\} \le \mathrm{e}^{-tLP + \frac{t^2L^2\E[E^2]}{2}} \label{appendixLemma1Eq9Block}.
\end{align}
Using~\eqref{appendixLemma1Eq1*Block}, \eqref{appendixLemma1Eq4Block},  \eqref{appendixLemma1Eq7Block} and~\eqref{appendixLemma1Eq8Block}, we obtain
\begin{align}
\Pr\left\{\tau(m)< \infty\right\}= \Pr\left\{\hat B_{\infty}< 0\right\}  \le \mathrm{e}^{-m\big(tLP - \frac{t^2L^2\E[E^2]}{2}\big)}. \label{appendixLemma1Eq9Block}
\end{align}
Combining~\eqref{appendixLemma1Eq0Block} and~\eqref{appendixLemma1Eq9Block}, we have
\begin{align}
\Pr\left\{\left.|\mathcal{Q}^{(n)}(w)| \ge L\gamma+1\right| W=w\right\} &\le \mathrm{e}^{-(m+\gamma)\big(t LP - \frac{t^2 L^2 \E[E^2]}{2}\big)}. \label{appendixLemma1Eq10Block}
\end{align}
In order to obtain an upper bound on the RHS of~\eqref{appendixLemma1Eq10Block}, we define $\alpha$ and $\beta$ as in~\eqref{defAlphaBlock} and~\eqref{defBetaBlock} respectively and use the following two facts due to~\eqref{sufficientlySmallTBlock} and~\eqref{defS}:
\begin{align}
t\le \frac{2\rho P}{L\E[E^2]+3S^2}=\alpha \label{tUpperBoundBlock}
\end{align}
and hence
\begin{align}
t\ge  \frac{2\rho P}{L\E[E^2]+3S^2(1+63\alpha S)}\ge  \beta. \label{tLowerBoundBlock}
\end{align}
Combining~\eqref{appendixLemma1Eq10Block}, \eqref{tUpperBoundBlock} and~\eqref{tLowerBoundBlock}, we conclude that
\eqref{stLemmaMismatchBlock} holds for any natural number~$\gamma$. It remains to show that~\eqref{stLemmaMismatchBlock} also holds if $\gamma$ is an arbitrary positive real number, which holds true due to the simple fact that
$\Pr\left\{\left.|\mathcal{Q}^{(n)}(w)| \ge L\gamma+1\right|W=w\right\} = \Pr\left\{|\mathcal{Q}^{(n)}(w)| \ge \lceil L\gamma+1\rceil\right\} $ for any $\gamma\in\mathbb{R}_+$.

\section{Proofs of Lemma~\ref{lemmaFeinstein} and Lemma~\ref{lemmaFeinsteinBlock}} \label{appendixA+Block}
Since save-and-transmit defined in Section~\ref{subsecSaveTransmit} is a special case of blockwise save-and-transmit defined in Section~\ref{subsecSaveTransmitBlock} with $L=1$ and Lemma~\ref{lemmaFeinstein} is a special case of Lemma~\ref{lemmaFeinsteinBlock} with $L=1$, it suffices to prove Lemma~\ref{lemmaFeinsteinBlock}.

Suppose we are given a blockwise save-and-transmit $(n, M)$-EH code.
Fix an~$L< n/m$, a $\gamma\ge 0$, a $\delta>0$ and an~$M\in \mathbb{N}$. We would like to obtain an upper bound on
\[
\Pr\left\{\left.\left\{\log\frac{p_{Y^n|X^n}(\tilde Y^n(1)|X^n(2))}{p_{Y^n}(\tilde Y^n(1))} > \log M + \delta\,\right\} \cap\left\{ |\mathcal{Q}^{(n)}(1)|<L\gamma+1\right\} \right|W=1\right\}
\]
 by a change-of-measure argument. To this end, we let $X^n=X^n(1)$, $\tilde X^n=\tilde X^n(1)$, $Y^n=Y^n(1)$ and $\tilde Y^n = \tilde Y^n(1)$ and use the definition of blockwise save-and-transmit in Section~\ref{subsecSaveTransmitBlock} and the definition of~$\mathcal{Q}^{(n)}(w)$ in~\eqref{defSetQ} to obtain
\begin{align}
p_{E^n, X^n, Y^n, \tilde X^n, \tilde Y^n, \mathcal{Q}^{(n)}(1)|W=1}&=\left(\prod_{k=1}^n p_{E_k,X_k,Y_k|W=1} p_{\tilde X_k | X_k, E^k, \tilde X^{k-1},W=1}p_{\tilde Y_k|\tilde X_k}\right)p_{\mathcal{Q}^{(n)}(1)|X^n, \tilde X^n, W=1} \label{appendixA+eq1}
\end{align}
where
\begin{align}
p_{\tilde Y_k|\tilde X_k}(\tilde y_k|\tilde x_k) \equiv q_{Y|X}(\tilde y_k|\tilde x_k), \label{equality2AppendixA+}
\end{align}
$p_{\tilde X_k | X_k, E^k, \tilde X^{k-1}, W=1}$ is some distribution readily determined by the encoding function~\eqref{defSavingEncodingBlock},
and
\begin{align*}
p_{\mathcal{Q}^{(n)}(1)|X^n, \tilde X^n, W=1} (\mathcal{A}|x^n, \tilde x^n)\equiv\begin{cases} 1& \text{if $\mathcal{A}=\{i\in\{m+1, m+2, \ldots, n\}|\,\tilde x_i\ne x_i\}$,}  \\
 0& \text{otherwise.}
\end{cases}  
\end{align*}
Using~\eqref{appendixA+eq1} and~\eqref{equality2AppendixA+}, we obtain
\begin{align*}
&p_{E^n, X^n, Y^n, \tilde X^n, \tilde Y^n, \mathcal{Q}^{(n)}(1)|W=1}(e^n, x^n, y^n, \tilde x^n, \tilde y^n, \mathcal{A}) \notag\\*
& \le \left(\prod_{k=1}^n p_{E_k, X_k, Y_k|W=1}(e_k, x_k, y_k)p_{\tilde X_k | X_k, E^k, \tilde X^{k-1},W=1}(\tilde x_k|x_k, e^k, \tilde x^{k-1})\right) \notag\\*
&\qquad \times \bigg(\prod_{k\in\mathcal{A}}q_{Y|X}(\tilde y_k | 0)\bigg)\bigg(\prod_{k\notin\mathcal{A}}q_{Y|X}(\tilde y_k|x_k)\bigg), 
\end{align*}
for each $(e^n, x^n, y^n, \tilde x^n, \tilde y^n)$ and each $\mathcal{A}\subseteq \{1, 2, \ldots, n\}$,
which implies by summing over $(e^n, x^n, y^n, \tilde x^n)$ that
\begin{align}
p_{\tilde Y^n, \mathcal{Q}^{(n)}(1)|W=1}(\tilde y^n, \mathcal{A})\le\left(\prod_{k\in\mathcal{A}} \mathcal{N}(\tilde y_k; 0, 1)\right)\left(\prod_{k\notin\mathcal{A}} \mathcal{N}(\tilde y_k; 0, S+1)\right) \label{appendixA+eq2}
\end{align}
 for all $(\tilde y^n, \mathcal{A})$.
 Consider the following chain of inequalities for each $\mathcal{A}\subseteq\{1, 2, \ldots, n\}$:
 \begin{align}
 &\Pr\left\{\left. \log\frac{p_{Y^n|X^n}(\tilde Y^n(1)|X^n(2))}{p_{Y^n}(\tilde Y^n(1))} > \log M + \delta\,\right| \mathcal{Q}^{(n)}(1)=\mathcal{A}, W=1 \right\} \notag\\*
&\quad = \int_{\mathbb{R}^n}\int_{\mathbb{R}^n} p_{X^n(2)}(x^n)p_{\tilde Y^n(1)| \mathcal{Q}^{(n)}(1), W=1}(\tilde y^n| \mathcal{A}) \notag\\*
&\qquad \quad\times \mathbf{1}\left\{ \frac{p_{\tilde Y^n(1),\mathcal{Q}^{(n)}(1)|W=1}(\tilde y^n,\mathcal{A})}{p_{Y^n}(\tilde y^n)} \times \frac{p_{Y^m}(\tilde y^m)\prod\limits_{k=m+1}^n p_{Y_k|X_k}(\tilde y_k|x_k)}{p_{\tilde Y^n(1),\mathcal{Q}^{(n)}(1)|W=1}(\tilde y^n,\mathcal{A})} > M \mathrm{e}^{ \delta}\right\} \mathrm{d} \tilde y^n \mathrm{d}x^n  \notag\\
&\quad \le \int_{\mathbb{R}^n}\int_{\mathbb{R}^n} p_{X^n(2)}(x^n)p_{\tilde Y^n(1)| \mathcal{Q}^{(n)}(1), W=1}(\tilde y^n| \mathcal{A}) \notag\\*
&\qquad \quad \times \mathbf{1}\left\{ (S+1)^{\frac{|\mathcal{A}|}{2}} \times \frac{p_{Y^m}(\tilde y^m)\prod\limits_{k=m+1}^n p_{Y_k|X_k}(\tilde y_k|x_k)}{p_{\tilde Y^n(1),\mathcal{Q}^{(n)}(1)|W=1}(\tilde y^n,\mathcal{A})} > M \mathrm{e}^{ \delta}\right\} \mathrm{d} \tilde y^n \mathrm{d}x^n \label{eq3AppendixA+StepA} \\
& \quad \le \frac{\mathrm{e}^{-\delta}}{M p_{\mathcal{Q}^{(n)}(1)|W=1}(\mathcal{A})}\,\times (S+1)^{\frac{|\mathcal{A}|}{2}} \notag\\*
&\qquad \times  \E\left[\frac{p_{Y^m}(\tilde Y^m(1))\prod\limits_{k=m+1}^n p_{Y_k|X_k}(\tilde Y_k(1)|X_k(2))}{p_{\tilde Y^n(1)|\mathcal{Q}^{(n)}(1)=\mathcal{A}, W=1 }(\tilde Y^n(1))}\right] \label{eq3AppendixA+StepB} \\
&\quad = \frac{\mathrm{e}^{-\delta}}{M p_{\mathcal{Q}^{(n)}(1)|W=1}(\mathcal{A})}\,\times (S+1)^{\frac{|\mathcal{A}|}{2}}   \label{eq3AppendixA+Block}
\end{align}
where
\begin{itemize}
\item \eqref{eq3AppendixA+StepA} is due to~\eqref{appendixA+eq2} and the fact that for all $y\in\mathbb{R}$,
$
\frac{\mathcal{N}(y; 0, 1)}{\mathcal{N}(y; 0, S+1)} \le \sqrt{S+1}
$;
\item \eqref{eq3AppendixA+StepB} follows from Markov's inequality where the expectation if evaluated with respect to the distribution ${p_{X^n(2)}p_{\tilde Y^n(1)|\mathcal{Q}^{(n)}(1)=\mathcal{A}, W=1}}$;
    \item \eqref{eq3AppendixA+Block} is due to simplifying the expectation term by first principles.
    \end{itemize}
Consequently,
\begin{align}
&\Pr\left\{\left.\left\{ \log\frac{p_{Y^n|X^n}(\tilde Y^n(1)|X^n(2))}{p_{Y^n}(\tilde Y^n(1))} > \log M + \delta\,\right\} \cap\left\{ |\mathcal{Q}^{(n)}(1)|<L\gamma+1\right\}\right|W=1 \right\} \notag\\*
&\quad \le \sum\limits_{\substack{\mathcal{A}\subseteq \{mL+1, mL+2, \ldots, n\}: \\ |\mathcal{A}|\le  L\gamma+1 }} \Pr\left\{\left. \log\frac{p_{Y^n|X^n}(\tilde Y^n(1)|X^n(2))}{p_{Y^n}(\tilde Y^n(1))} > \log M + \delta\,\right|  \mathcal{Q}^{(n)}(1)=\mathcal{A}, W=1\right\} \notag\\*
&\qquad \times \Pr\{ \mathcal{Q}^{(n)}(1)=\mathcal{A}|W=1\}\notag\\
&\quad \le \frac{\mathrm{e}^{-\delta}}{M}\times (S+1)^{\frac{L\gamma + 1}{2}}\times \big|\mathcal{A}\subseteq \{mL+1, mL+2, \ldots, n\}:  |\mathcal{A}|\le L\gamma+1 \big| \label{eq4AppendixA+Block}
\end{align}
where the last inequality is due to~\eqref{eq3AppendixA+Block}. Since the mismatched positions occur in blocks of~$L$ symbols except for the last block whose length is no larger than~$L$, we have
\begin{align}
\big|\mathcal{A}\subseteq \{mL+1, mL+2, \ldots, n\}:  |\mathcal{A}|\le L\gamma+1 \big|  &\le \sum\limits_{i=0}^{\lceil \gamma+1/L\rceil} \binom{\lceil(n-mL)/L\rceil}{i} \notag\\*
&\le \sum\limits_{i=0}^{\lceil \gamma+1/L\rceil}\left(\left\lceil\frac{n-mL}{L}\right\rceil\right)^i\notag \\
& \le \sum\limits_{i=0}^{\lceil \gamma+1/L\rceil}\left(n-mL\right)^i\notag \\
& \le \frac{\left(n-mL\right)^{\gamma+2}}{n-mL-1} \notag\\
&\le  2(n-mL)^{\gamma+1}. \label{eq5AppendixA+Block}
\end{align}
Combining~\eqref{eq4AppendixA+Block} and~\eqref{eq5AppendixA+Block}, we obtain~\eqref{stLemmaFeinsteinBlock}.

\section{Proofs of Corollary~\ref{corollaryThmMainResultAWGNsaving} and Corollary~\ref{corollaryThmMainResultAWGNsavingBlock}}\label{appendixDblock}
Since save-and-transmit defined in Section~\ref{subsecSaveTransmit} is a special case of blockwise save-and-transmit defined in Section~\ref{subsecSaveTransmitBlock} with $L=1$ and Corollary~\ref{corollaryThmMainResultAWGNsaving} is a special case of Corollary~\ref{corollaryThmMainResultAWGNsavingBlock} with $L=1$, it suffices to prove Corollary~\ref{corollaryThmMainResultAWGNsavingBlock}.

Fix an $\varepsilon\in(0,1/2)$, and fix any $\varepsilon_1>0$ and $\varepsilon_2>0$ such that $\varepsilon_1+\varepsilon_2= \varepsilon$. Define $\rho$, $S$, $\alpha$, $\beta$ and~$m$ as in~\eqref{defRhoSavingBlock}, \eqref{defS}, \eqref{defAlphaBlock}, \eqref{defBetaBlock} and~\eqref{defmSavingBlock} respectively, and define $\gamma(\varepsilon_2)=0$ as in~\eqref{defGammaBlock}. To simplify notation, we do not explicitly specify the dependence on~$n$ for $\rho$, $S$, $\alpha$, $\beta$ and~$m$. Let $p_X= \mathcal{N}(x;0, S)$ and let $p_Y=\mathcal{N}(y; 0, S+1)$ be the marginal distribution of~$p_Xq_{Y|X}$, and let $\sigma^2$ and $T$ denote the variance and the third absolute moment of
$
 \log\frac{q_{Y|X}(Y|X)}{p_{Y}(Y)}
$
respectively. Fix any sufficiently large~$n$ and any $L< n$ such that $\rho\in(0,1)$, \eqref{assumptionThm} and~\eqref{sufficientlyLargeN} simultaneously hold. Then,
Theorem~\ref{thmMainResultContAWGNsavingBlock} implies that there exists a blockwise save-and-transmit $(n, M, \varepsilon)$-EH code which satisfies~\eqref{st1ThmMainResultAWGNblock} and~\eqref{st3ThmMainResultAWGNblock}. We would like to show that~\eqref{st1CorollaryMainResultAWGNsavingBlock} holds for the blockwise save-and-transmit code by obtaining a lower bound on the RHS of~\eqref{st1ThmMainResultAWGNblock}. To this end, we fix a sufficiently large~$n$ such that~\eqref{st1ThmMainResultAWGNblock} holds for the blockwise save-and-transmit code.
By construction, we have $S=P(1-\rho )$, $\rho =\Theta\Big(\sqrt{\frac{L}{n}}\Big)$ and $m=\Theta\big(\sqrt{\frac{n}{L}}\big)$, and we use Taylor's theorem to conclude that there exist some $\kappa_1>0$ and $\kappa_2>0$ which do not depend on~$n$ such that
 \begin{align}
\frac{1}{2}\log (1+S) &\ge \frac{1}{2}\left(\log(1+P)-\frac{\rho  P}{1+P}-\frac{\kappa_1 L}{n}\right) \label{eq1CorollaryMainResultAWGNsavingBlock}
\end{align}
and
\begin{align} \Phi^{-1}(\varepsilon_1) - \frac{\kappa_2}{\sqrt{n}}\le \Phi^{-1}\left(\varepsilon_1 -\frac{T}{\sigma^3\sqrt{n_m}}-\frac{4}{\sqrt{n_m}}\right)<0. \label{eq2CorollaryMainResultAWGNsavingBlock}
\end{align}
Combining~\eqref{st1ThmMainResultAWGNblock}, \eqref{eq1CorollaryMainResultAWGNsavingBlock} and~\eqref{eq2CorollaryMainResultAWGNsavingBlock} and using the facts that $\rho =\Theta\Big(\sqrt{\frac{L}{n}}\Big)$, $m=\Theta\big(\sqrt{\frac{n}{L}}\big)$ and
\begin{align*}
\sqrt{n\sigma^2}- \sqrt{mL\sigma^2}\le\sqrt{(n-mL) \sigma^2},
\end{align*}
we obtain
\begin{align}
\frac{1}{n}\log M
&\ge \frac{n-mL}{2n}\log (1+S) + \frac{\sqrt{(n-mL) \sigma^2}}{n}\,\Phi^{-1}\left(\varepsilon_1 -\frac{T}{\sigma^3\sqrt{n_m}}-\frac{4}{\sqrt{n_m}}\right) - \frac{\log \sqrt{n}+1}{n}\notag\\*
&\qquad-\frac{1}{n}\left(L\Big(2S\log 2+\frac{1}{2}\log(1+S)\Big) + (8S+1)\log n_m\right) \notag\\*
&\ge \frac{1}{2}\log (1+P) -\frac{\rho  P}{2(1+P)} -  \frac{mL}{2n}\log (1+P)   + \sqrt{\frac{\sigma^2}{n}}\,\Phi^{-1}(\varepsilon_1)-\kappa_3 \max\left\{\frac{ L^{1/4}}{n^{3/4}}, \frac{L}{n}\right\}  \label{eq3CorollaryMainResultAWGNsavingBlock}
\end{align}
for some $\kappa_3>0$ which does not depend on~$n$. In order to bound the second term on the RHS of~\eqref{eq3CorollaryMainResultAWGNsavingBlock}, we obtain from the definition  of~$\rho$ in~\eqref{defRhoSavingBlock} that
\begin{align}
\frac{\rho  P}{2(1+P)} = \frac{\sqrt{(L\E[E^2]+3P^2)\log(1+P)\log\frac{1}{\varepsilon_2}}}{2\sqrt{2nP(1+P)}}= \Theta\left(\sqrt{\frac{L}{n}}\right). \label{eq4CorollaryMainResultAWGNsavingBlock}
\end{align}
In order to bound the third term on the RHS of~\eqref{eq3CorollaryMainResultAWGNsavingBlock}, we obtain the following bounds by the definition of~$\alpha=\Theta\Big(\sqrt{\frac{1}{Ln}}\Big)$ in~\eqref{defAlphaBlock}, the definition of~$\beta$ in~\eqref{defBetaBlock} and the definition of~$\gamma(\varepsilon_2)$ in~\eqref{defGammaBlock} where $\kappa_4$ and $\kappa_5$ are some positive constants that do not depend on~$n$:
\begin{align}
\alpha
&\ge \frac{2\rho  P}{L\E[E^2]+3P^2} \label{alphaBoundSavingBlock}
 \end{align}
where~\eqref{alphaBoundSavingBlock} follows from the definition of $\rho$ in~\eqref{defRhoSavingBlock};
  \begin{align}
\beta
&\ge \frac{\alpha}{1+63\alpha P} \notag\\
&\ge \frac{2\rho  P}{L\E[E^2]+3P^2} - \frac{\kappa_4}{Ln} \label{betaBoundSavingBlock}
 \end{align}
where~\eqref{betaBoundSavingBlock} is due to~\eqref{alphaBoundSavingBlock};
\begin{align}
m &= \left\lceil\frac{\log\frac{1}{\varepsilon_2}}{LP\beta+\frac{L^2\alpha^2 \E[E^2]}{2}}\right\rceil  \notag\\*
&\le \frac{\log\frac{1}{\varepsilon_2}}{LP\beta} +1 \notag\\
&=\frac{(L\E[E^2]+3P^2)\log\frac{1}{\varepsilon_2}}{2\rho  L P^2}\times\frac{\frac{2\rho  P}{L\E[E^2]+3P^2} }{\beta} +1 \notag\\*
&\le \frac{(L\E[E^2]+3P^2)\log\frac{1}{\varepsilon_2}}{2\rho L P^2} +\frac{\kappa_5}{L}
\label{mBoundSavingBlock}
\end{align}
where~\eqref{mBoundSavingBlock} is due to~\eqref{betaBoundSavingBlock} and the definition of $\rho$ in~\eqref{defRhoSavingBlock}. Using~\eqref{mBoundSavingBlock} and the definition of $\rho$ in~\eqref{defRhoSavingBlock}, we have
\begin{align}
m \le \frac{1}{L}\sqrt{\frac{n(L\E[E^2]+3P^2)\log\frac{1}{\varepsilon_2}}{2P(P+1)\log(1+P) }}  + \frac{\kappa_5}{L} .  \label{eq5CorollaryMainResultAWGNsavingBlock}
\end{align}
Combining~\eqref{eq3CorollaryMainResultAWGNsavingBlock}, \eqref{eq4CorollaryMainResultAWGNsavingBlock} and~\eqref{eq5CorollaryMainResultAWGNsavingBlock}, we conclude that \eqref{st1CorollaryMainResultAWGNsavingBlock} holds for any sufficiently large~$n$ where $\kappa>0$ is some constant which does not depend on~$n$.

In addition, \eqref{st2CorollaryMainResultAWGNsavingBlock} follows from the following inequality due to~\eqref{st3ThmMainResultAWGNblock}, the definition of $\mathcal{Q}^{(n)}(w)$ in~\eqref{defSetQ} and our choice for $\gamma(\varepsilon_2)$ that $\gamma(\varepsilon_2)=0$:
\begin{align*}
\Pr\left\{\bigcup_{k=mL+1}^n\left\{\sum_{i=1}^k E_i < \sum_{i=m+1}^k X_i^2  \right\} \right\} & = \Pr\left\{|\mathcal{Q}^{(n)}(W)|\ge 1   \right\}\\
&=\Pr\left\{|\mathcal{Q}^{(n)}(W)|\ge L\gamma(\varepsilon_2)+ 1  \right\}\\*
& \le \varepsilon_2.
\end{align*}
\section{Proofs of Corollary~\ref{corollaryThmMainResultAWGN} and Corollary~\ref{corollaryThmMainResultAWGNblock}} \label{appendixBblock}
Since best-effort defined in Section~\ref{subsecBestEffort} is a special case of blockwise best-effort defined in Section~\ref{subsecBestEffortBlock} with $L=1$ and Corollary~\ref{corollaryThmMainResultAWGN} is a special case of Corollary~\ref{corollaryThmMainResultAWGNblock} with $L=1$, it suffices to prove Corollary~\ref{corollaryThmMainResultAWGNblock}.

Fix an $\varepsilon\in(0,1/2)$, and fix any $\varepsilon_1>0$ and $\varepsilon_2>0$ such that $\varepsilon_1+\varepsilon_2= \varepsilon$. Define $\rho$, $S$, $\alpha$ and $\beta$ as in~\eqref{defRhoBlock}, \eqref{defS}, \eqref{defAlphaBlock} and \eqref{defBetaBlock} respectively. In addition, let $m\triangleq 0$ and define
$\gamma(\varepsilon_2)$ as in~\eqref{defGammaBestEffortBlock}. To simplify notation, we do not explicitly specify the dependence on~$n$ for $\rho$, $S$, $\alpha$, $\beta$ and~$\gamma(\varepsilon_2)$. Let $p_X= \mathcal{N}(x;0, S)$ and let $p_Y=\mathcal{N}(y; 0, S+1)$ be the marginal distribution of~$p_Xq_{Y|X}$, and let $\sigma^2$ and $T$ denote the variance and the third absolute moment of $\log\frac{q_{Y|X}(Y|X)}{p_{Y}(Y)}$
respectively. Fix any sufficiently large~$n$ such that $\rho\in(0,1)$, $\varepsilon_1 -\frac{T}{\sigma^3\sqrt{n}}-\frac{4}{\sqrt{n}}>0$ and \eqref{sufficientlyLargeN} simultaneously hold. Then, Theorem~\ref{thmMainResultContAWGNsavingBlock} implies that there exists a blockwise best-effort $(n, M, \varepsilon)$-EH code which satisfies
\begin{align}
\log M &\ge \frac{n}{2}\log (1+S) + \sqrt{n \sigma^2}\,\Phi^{-1}\left(\varepsilon_1 -\frac{T}{\sigma^3\sqrt{n}}-\frac{4}{\sqrt{n}}\right) \notag\\*
&\qquad-\left(\lambda_1 L + \lambda_2 \log n\right) (\gamma(\varepsilon_2)+1) - \log \sqrt{n} -1 \label{stThmMainResultAWGNinProofBlock}
\end{align}
and~\eqref{st3ThmMainResultAWGNblock}
where $\lambda_1$ and $\lambda_2$ are as defined in~\eqref{defLambda1} and~\eqref{defLambda2} respectively. In the rest of the proof, we will derive~\eqref{st1CorollaryMainResultAWGNblock} from~\eqref{stThmMainResultAWGNinProofBlock}.
By construction, we have $S=P(1-\rho)$ and $\rho=\Theta\Big(\sqrt{\frac{\log n}{n}}\Big)$, and we use Taylor's theorem to conclude that there exist some $\kappa_1>0$ and $\kappa_2>0$ which do not depend on~$n$ such that
 \begin{align}
\frac{1}{2}\log (1+S) &\ge \frac{1}{2}\left(\log(1+P)-\frac{\rho P}{1+P}-\frac{\kappa_1 \log n}{n}\right) \label{eq1CorollaryMainResultAWGNblock}
\end{align}
and
\begin{align}
\Phi^{-1}\left(\varepsilon_1 -\frac{T}{\sigma^3\sqrt{n}}-\frac{4}{\sqrt{n}}\right) \ge \Phi^{-1}(\varepsilon_1) - \frac{\kappa_2}{\sqrt{n}}. \label{eq2CorollaryMainResultAWGNblock}
\end{align}
Combining~\eqref{stThmMainResultAWGNinProofBlock}, \eqref{eq1CorollaryMainResultAWGNblock} and~\eqref{eq2CorollaryMainResultAWGNblock}, we obtain
\begin{align}
\frac{1}{n}\log M
& \ge  \frac{1}{2}\log (1+P) -\frac{\rho  P}{2(1+P)} -\frac{(\lambda_1 L + \lambda_2 \log n) (\gamma(\varepsilon_2)+1)}{n}    - \sqrt{\frac{\sigma^2}{n}}\,\Phi^{-1}(\varepsilon_1) \notag\\*
&\qquad - \frac{\kappa_2\sqrt{\sigma^2}}{n}- \frac{\log \sqrt{n}+1}{n}-\frac{\kappa_1 \log n}{2n}\notag\\*
&\ge \frac{1}{2}\log (1+P) -\frac{\rho  P}{2(1+P)} -\frac{(\lambda_1 L + \lambda_2 \log n)\gamma(\varepsilon_2)}{n}    - \sqrt{\frac{\sigma^2}{n}}\,\Phi^{-1}(\varepsilon_1)-\frac{\kappa_3 \max\{\log n,L\}}{n}  \label{eq3CorollaryMainResultAWGNblock}
\end{align}
for some $\kappa_3>0$ which does not depend on~$n$.
 In order to bound the second term on the RHS of~\eqref{eq3CorollaryMainResultAWGNblock}, we obtain from the definition of~$\rho$ in~\eqref{defRhoBlock} that
\begin{align}
\frac{\rho P}{2(1+P)} = \frac{1}{2}\sqrt{\frac{(\lambda_1 L + \lambda_2 \log n)(L\E[E^2]+3P^2)\log\frac{1}{\varepsilon_2}}{LP(P+1)n}}= \Theta\left(\sqrt{\frac{\max\{\log n, L\}}{n}}\right). \label{eq4CorollaryMainResultAWGNblock}
\end{align}
In order to bound the third term on the RHS of~\eqref{eq3CorollaryMainResultAWGNblock}, we obtain the following bounds by the definition of~$\alpha = \Theta\Big(\sqrt{\frac{\log n}{n}}\Big)$ in~\eqref{defAlphaBlock}, the definition of~$\beta$ in~\eqref{defBetaBlock} and the definition of~$\gamma(\varepsilon_2)$ in~\eqref{defGammaBlock} where $\kappa_4$ and $\kappa_5$ are some positive constants that do not depend on~$n$:
\begin{align}
\alpha
&\ge \frac{2\rho  P}{L\E[E^2]+3P^2} \label{alphaBoundBlock}
 \end{align}
where~\eqref{alphaBoundBlock} follows from the definition of $\rho$ in~\eqref{defRhoBlock};
  \begin{align}
\beta
&\ge \frac{\alpha}{1+63\alpha P} \notag\\
&\ge \frac{2\rho  P}{L\E[E^2]+3P^2} - \frac{\kappa_4\max\{\log n, L\}}{L^2 n} \label{betaBoundBlock}
 \end{align}
where~\eqref{betaBoundBlock} is due to~\eqref{alphaBoundBlock};
\begin{align}
\gamma(\varepsilon_2) &= \frac{\log\frac{1}{\varepsilon_2}}{PL\beta+\frac{L^2\alpha^2 \E[E^2]}{2}}  \notag\\
&\le \frac{\log\frac{1}{\varepsilon_2}}{PL\beta} \notag\\
&=\frac{(L\E[E^2]+3P^2)\log\frac{1}{\varepsilon_2}}{2\rho  LP^2}\times\frac{\frac{2\rho  P}{L\E[E^2]+3P^2} }{\beta}  \notag\\
&\le \frac{(L\E[E^2]+3P^2)\log\frac{1}{\varepsilon_2}}{2\rho L P^2} +\frac{\kappa_5}{L} \label{gammaBoundBlock}
\end{align}
where~\eqref{gammaBoundBlock} is due to~\eqref{betaBoundBlock} and the definition of $\rho$ in~\eqref{defRhoBlock}.
 Using~\eqref{gammaBoundBlock} and the definition of $\rho$ in~\eqref{defRhoBlock}, we have
\begin{align}
\gamma(\varepsilon_2) \le \frac{1}{2}\sqrt{\frac{n(L\E[E^2]+3P^2)\log\frac{1}{\varepsilon_2}}{LP(P+1)(\lambda_1 L + \lambda_2 \log n)}} + \frac{\kappa_5}{L}.   \label{eq5CorollaryMainResultAWGNblock}
\end{align}
Combining~\eqref{eq3CorollaryMainResultAWGNblock}, \eqref{eq4CorollaryMainResultAWGNblock} and~\eqref{eq5CorollaryMainResultAWGNblock}, we conclude that \eqref{st1CorollaryMainResultAWGNblock} holds for any sufficiently large~$n$ where $\kappa>0$ is some constant which does not depend on~$n$.

\section*{Acknowledgment}
We would like to thank the Associate Editor and the anonymous reviewers for their valuable comments which helped us improve the presentation of this work.



%
%
%


\end{document}